# ElemeNet: Multiscale Molecular Machine Learning with Uncertainty Quantification Across the Periodic Table

Jacob W. Toney[1,2,#], Samir Darouich[1,3,4,#], Yiran Wang[1,2], Aaron G. Garrison[1], Johannes Kästner[3], and Heather J. Kulik[1,2,5,*]

[1]*Department of Chemical Engineering, Massachusetts Institute of Technology, Cambridge, MA 02139, USA*
[2]*Center for Computational Science and Engineering, Massachusetts Institute of Technology, Cambridge, MA 02139, USA*
[3]*Institute for Theoretical Chemistry, University of Stuttgart, Germany*
[4]*Institute for Artificial Intelligence, University of Stuttgart, Germany*
[5]*Department of Chemistry, Massachusetts Institute of Technology, Cambridge, MA 02139, USA*
#These authors contributed equally
*corresponding author email: hjkulik@mit.edu

ABSTRACT: Advances in deep learning architectures and representations have enabled ML-driven chemical property prediction, but state-of-the-art (SOTA) models have remained largely confined to independent codebases and lack support for diverse chemical species. This work introduces ElemeNet, a unified, general-purpose software package for molecular machine learning. The ElemeNet software package enables the training of advanced ML models for diverse properties and datasets with an enlarged range of elemental compositions. We define molecular representations compatible with elements 1–100, supporting diverse organometallic and biological systems in addition to organic chemistry already well-served by the Chemprop ML toolkit. As well as more common atom-, bond-, and molecule-level predictions, we introduce moiety predictions. We also natively define optional conditioning on charge and spin states. Advanced E(3)-equivariant and transformer architectures are supported, as well as classical 2D models, with all classes including built-in uncertainty quantification through deterministic and statistical measures. We benchmark our protocols for ML model training against representative datasets from organic, inorganic, coordination, and biological chemistry, achieving competitive and SOTA performance relative to literature baselines and favorable scaling to millions of molecules. The entire workflow is exposed through a concise command-line interface, lowering the barrier to entry for non-expert users. We anticipate ElemeNet will empower non-computational researchers to leverage modern deep learning methods across the chemical and physical sciences.

## 1. Introduction.

Machine learning (ML) has become a central tool in computational chemistry, reshaping molecular property prediction and chemical space exploration. Early heuristic structure–property models aimed at capturing linear free-energy relationships[1-4] were largely supplanted by multivariate regression for navigating more diverse chemical spaces.[5-13] Deep learning further expanded the expressive capacity of models, allowing task-relevant features and relationships to be learned directly from data rather than through handcrafted features.[14-16] This progression has been mirrored by an evolution in molecular representations, from fixed-length encodings[17] and string-based representations[18] to two- and three-dimensional molecular graphs.[19-25] Model architectures have advanced in tandem: Gaussian processes and multilayer perceptrons (MLPs) have frequently been replaced by graph neural networks (GNNs), which operate directly on molecular graphs, as datasets became larger.[26-29] The introduction of symmetry-equivariance and attention have since defined state-of-the-art (SOTA) architectures.[29-34] Structure–property prediction GNN models are key players in high-throughput virtual screening for molecular discovery and optimization, and GNNs serve as the foundation of modern generative models and machine-learning interatomic potentials (MLIPs).[35-46] These representations and architectures have leveraged the emergence of large quantum-chemical databases, to bring deep learning to the forefront of computational chemistry.[47-50]

These advances have enabled the development and training of increasingly accurate models for chemical property prediction. In practice, however, these developments are largely confined to specialized codebases that are difficult or impossible to generalize beyond the chemical contexts for which they were designed. The slow pace at which technical advances reach general-purpose software has limited their adoption. The Chemprop package represented a significant step toward democratizing ML for organic chemistry.[51,52] Wrapped in a convenient and accessible framework, the codebase allows chemists without ML expertise to train performant GNN models on their own datasets. Nevertheless, the Chemprop architecture carries several important limitations, foremost among them its restriction to 2D models trained on SMILES inputs. While SMILES strings encode molecular connectivity, they do not natively represent stereochemistry, limiting the expressive capacity of all models. SMILES are further known to inadequately describe systems with complex bonding, including organometallic complexes, heavy elements, and biological macromolecules that are central to catalysis, magnetism, and drug discovery.[20,23,53-56] Equivariant and attention-based frameworks employed by SOTA models are also not supported. Extending molecular machine learning workflows to the full periodic table requires extensions beyond existing general-purpose tools.

The electronic structure of molecules is a function of the 3D coordinates of all atoms in a system, limiting the ability of 2D methods to reproduce high-accuracy *ab initio* or experimental reference data. Although 2D representations retain clear utility, such as when 3D information is unavailable or in 2D-to-3D learning tasks, models capable of capturing complex bonding patterns are ultimately required for property prediction spanning the periodic table. Beyond atomic coordinates, charge and spin are necessary to fully specify a quantum-chemical system and must also be encoded. An ML codebase for general-purpose chemistry must also keep pace with increasing dataset sizes, scaling efficiently to millions of molecules. Finally, uncertainty quantification is a critical but frequently overlooked component of molecular ML tasks. Reliable uncertainty estimates are valuable not only for knowing when an individual prediction can be trusted but are a strict prerequisite for active learning campaigns.[57-63] Methods based on latent-space distance have proven effective but are limited to the topology of the latent manifold (i.e., the quality of the learned representation).[64-70] Fully Bayesian ensemble-based approaches are ideal but incur substantial computational cost.[71-75] Recently introduced shallow, last-layer ensembles address this through a quasi-Bayesian approximation to a deep ensemble.[76,77]

In this work, we introduce ElemeNet, a modular software package for molecular machine learning across the periodic table. ElemeNet accepts diverse molecular input types spanning both 2D and 3D representations, supporting single- and multitask property prediction at the levels of atoms, bonds, and molecules. It additionally enables novel subgraph-level targets for direct prediction on chemical moieties. Graph-level embeddings further allow prediction of molecular properties as explicit functions of charge and spin, invaluable for modern efforts to integrate ML and electronic structure theory. Our codebase implements advanced GNN architectures, including 3D equivariant encoders and transformers absent from peer end-to-end packages, while retaining support for classical 2D models with MLP readout. ElemeNet scales well to datasets containing millions of molecules and generates well-calibrated uncertainty estimates for all prediction tasks. Benchmarks across tasks in organic, inorganic, biological, and coordination chemistry reveal ElemeNet to be competitive with and to frequently outperform representative literature baselines. We anticipate that ElemeNet will serve as a practical platform for training performant deep learning models across chemical domains and datasets, democratizing machine learning for chemistry.

## 2. Computational Details.

### 2a. Code Dependencies.

ElemeNet is constructed using several existing software packages as dependencies. PyTorch[78] version 2.8.0 and PyTorch Geometric[79] version 2.6.1 are used for graph, dataloader, and model, construction. Hyperparameter optimization is performed using Optuna[80] version 4.5.0. RDKit[81] version 2025.03.5 is used for parsing SMILES strings, while scikit-learn[82] version 1.7.2 is used for defining data splits. Experiments were run across different Nvidia Tesla V100, L40S, and H200 GPUs as available.

**2b. Model Baselines.**

Model evaluations performed on the QM9, tmQMg, pydentate, and GEMS datasets were implemented to maintain consistency with corresponding literature baselines. The QM9 dataset includes 133,885 small organic molecules containing up to nine heavy atoms (C, N, O, F) computed at the DFT level.[47] For QM9 experiments, all models were trained using the data splits from Heid et al.[51] 2D models were trained using SMILES string inputs both with and without optional RDKit features, with the results of the best-performing models reported. For all 2D models, deep ensembles of 5 models with identical configurations were trained and their predictions averaged, using the same ensembling procedure as Heid et al. 3D models were trained using xyz inputs, with no deep ensembling. Implicit hydrogens were used for both 2D and 3D models. All models were hyperparameter optimized for 25 trials of 100 epochs each.

The tmQMg dataset contains 60,799 neutral and singly charged mononuclear transition metal complexes from the Cambridge Structural Database[83] with geometries optimized and properties calculated with DFT.[48,84] The tmQMg experiments used the same splits from Kneiding et al.[84], omitting the same outliers identified and excluded in the original tmQMg work. 2D models were trained using mol2 inputs, while 3D models were trained using xyz inputs, and both types of models used explicit hydrogens. Molecular charge as reported in tmQMg is used as an input to the charge-spin neural network. All 2D models were hyperparameter optimized for 25 trials of 100 epochs each. All 3D models with MLP readout used the optimal hyperparameters identified by the corresponding 2D model. 3D models with transformer readout used default hyperparameters.

The pydentate baselines trained on 66,355 ligands extracted from the Cambridge Structural Database[83] used the same structures and splits from Toney et al.[85] Only 2D GNN models were considered. Models trained to predict coordinating atoms and coordination number used SMILES inputs with implicit hydrogens, omitting the 66 of 66,355 structures for which SMILES strings could not be parsed. Models trained to predict bond lengths and metal–ligand bond lengths used mol2 inputs and explicit hydrogens. All models were hyperparameter optimized for 25 trials of 100 epochs each.

The GEMS dataset includes 2,713,986 charged, off-equilibrium fragments of biological molecules with DFT-calculated properties.[86] We developed GEMS baselines using random splits because there were no prior data splits or benchmarks on this dataset. Only 3D GNN models with transformer readout were considered. Models were trained from xyz inputs with explicit hydrogen atoms using default hyperparameters. Molecular charge as reported in the GEMS dataset is used as an input to the charge-spin neural network.

For all baselines obtained in this work across all datasets and tasks considered, models were initially trained for 500 epochs with early stopping based on validation loss. In the case that the minimum validation loss was observed within the final 100 epochs of training, models were finetuned for up to 1000 epochs for 2D models and 2000 epochs for 3D models, again with early stopping. 2D models were trained on a single GPU, 3D models were trained across 4 GPUs.

## 3. Results and Discussion.

### 3a. Model Architectures and Representations.

We constructed the ElemeNet code to support molecular (i.e., finite) systems containing elements spanning the periodic table, with explicit support for atomic numbers 1–100 (i.e., hydrogen to fermium). The code supports both 2D and 3D graph learning, each with a unique method of molecular graph construction (Figure 1). This flexibility is meant to accommodate unique modeling requirements and data availability among chemical domains (e.g., organic and inorganic chemistry). The 2D representations (i.e., derived from topological information in chemical bonds) lack any 3D information, as coordinates are used only for bond interpretation and subsequently discarded. As such, these molecular graphs are less computationally expensive than their 3D counterparts but lack any encoding of stereochemistry, making them well-suited to target properties primarily determined by 2D interactions (e.g., physicochemical properties[87]). In the case of learning tasks where 3D information is necessary (e.g., interatomic potentials[88], electronic structure properties[89]), molecular graphs are constructed from 3D coordinates. In this case, we define edges pairwise between all atoms within a tunable cutoff radius, with only the 3D distance between atoms being used as edge features to impose E(3) equivariance[32]. For both 2D and 3D input types, the code featurizes the resulting molecular graphs and passes them to the downstream encoder without further preprocessing of connectivity.

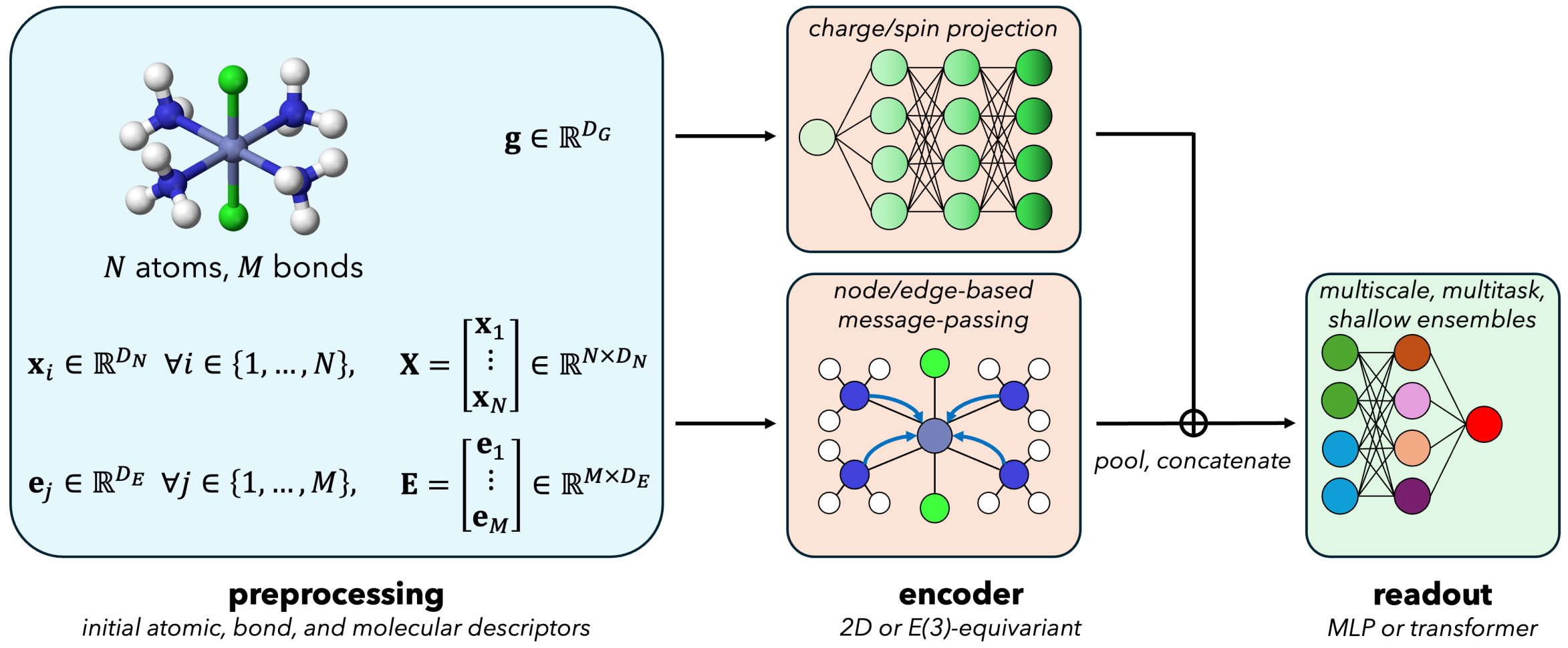

**Figure 1.** Block diagram representation of the ElemeNet architecture. Initial node, edge, and graph representations are constructed from a given molecular input (left). Node and edge features are used to learn latent embeddings through message-passing in a 2D or equivariant 3D GNN (bottom middle), while graph features of charge and spin are projected through a dedicated neural network (top middle). Learned node representations are pooled and concatenated to graph embeddings before being processed by a readout MLP or transformer for property prediction (right). While graph-level property prediction is depicted for representative purposes, ElemeNet also supports multitask and multiscale targets (Section 3b) with ensemble-based uncertainty quantification (Section 3c).

Existing graph neural network frameworks for chemistry typically rely on initial representations tailored to main group organic chemistry (e.g., hybridization, covalent bond orders), which are less well defined for complex molecules (e.g., with noncovalent interactions or Jahn–Teller distortion). To define chemical priors that are broadly balanced across elements in the periodic table, our 2D and 3D models both utilize an initial node representation comprised of key tabulated atomic properties (Supporting Information Table S1). Specifically, this feature vector includes continuous features (i.e., atomic mass, covalent radius, van der Waals radius, electronegativity, polarizability, ionization potential, and electronegativity) that are represented as scaled values. We also add categorical features (i.e., group, period, and number of valence electrons in *s*, *p*, *d*, and *f* atomic orbitals) that are one-hot encoded based on values from a lookup table. Discrete features (i.e., coordination number and number of neighboring hydrogen atoms) are provided directly and scaled accordingly (Supporting Information Table S1). The code updates the resulting node feature vector through message-passing operations to achieve a learned representation, which is subsequently passed through a readout decoder for property prediction. Edge features are provided that encode key properties of covalent bond type (i.e., single, double, triple, or aromatic) as encoded by the input molecular graph, an estimated bond length (approximated by the sum of covalent radii between bonded atoms), and a Boolean label indicating whether the bond includes a metal (Supporting Information Table S2 and Text S1). Graph-level representations encoding the charge and spin state of a molecule are also included to enable development of models with the ability to capture changes in electronic structure (Figure 1). These features are each projected through a dedicated charge-spin neural network and concatenated to the learned node representations (Supporting Information Text S2 and S3).

While we observe strong performance with our default initial representation scheme, ElemeNet also supports augmenting this representation with additional features. Additional user-provided features may be provided at the node, edge, or graph level. In the case that these features include charge and spin information, these graph-level attributes are updated from their default values (i.e., assuming neutral atoms and low-spin states). Optional auxiliary chemical information is available from bulk material properties and semiempirical quantum chemical calculations[90] (Supporting Information Tables S3 and S4). When enabled, these additional features are concatenated to the corresponding initial atom, bond, and molecule-level representations. ElemeNet also supports the use of implicit hydrogen atoms, where all nodes corresponding to hydrogen atoms are not featurized and are ignored by the network, while the representations of all remaining heavy atoms are updated with an additional attribute encoding the number of attached hydrogens. This reduces the computational cost of model training at the expense of expressivity. As such, we advise against the use of implicit hydrogens for learning tasks with relevant hydrogen chemistry for which explicit encoding is warranted (e.g., hydrogen bonding, agostic interactions, and off-equilibrium structures). Once generated, the initial node, edge, and graph representations are transformed to learned embeddings and target predictions based on the specified model architecture.

ElemeNet adopts a modular architecture in which a learnable graph neural network encoder is used to extract latent embeddings from initial node, edge, and graph representations. These latent embeddings are subsequently passed to a task-specific readout module for property prediction. Supported encoder architectures are 2D graph neural networks (GNNs) and 3D equivariant graph neural networks (EGNNs), while the readout module may be a multilayer perceptron (MLP) or a transformer. Our 2D GNN encoders support a range of architectures, including the widely used directed message-passing algorithm introduced by Gilmer and coworkers[91] and implemented in several molecular machine learning workflows.[32,51,52,84] We also provide support for graph convolution[92,93], graph isomorphism[94], and graph attention networks[95], which are absent from existing codebases[51,52]. All GNN layers operate on input node features, $\mathbf{x}_i$, and, optionally, edge features, $\mathbf{e}_{ij}$ to compute learned embeddings, $\mathbf{x}_i^{(h)}$:

$$\mathbf{x}_i^{(h+1)} = f\left(\mathbf{x}_i^{(h)}, \mathbf{x}_{j\in\mathcal{N}(i)}^{(h)}, \mathbf{e}_{ij\in\mathcal{N}(i)}, \boldsymbol{\theta}\right) \quad (1)$$

where $\mathbf{x}_i^{(h)}$ is the initial representation for node $i$ at layer $h$, $\mathcal{N}(i)$ is the set of all nodes that share an edge with $i$, $\mathbf{e}_{ij}$ is the feature vector describing the edge between nodes $i$ and $j$, and $\boldsymbol{\theta}$ is a set of learnable parameters characterizing the update step. The functional form of the node embedding update will vary depending on the convolution operator selected (eqn. 1).[91-95] Repeating equation 1 for all $N$ atoms constitutes a single message-passing layer. This process is then repeated for each of the $H$ hidden layers in the graph neural network encoder, resulting in final learned node embeddings $\mathbf{X}^{(H)}$. In the case of graph-level target properties, the learned node embeddings are pooled and concatenated to any available graph-level features $\mathbf{g}$, resulting in a graph-level representation $\mathbf{g}'$:

$$\mathbf{g}' = \text{concat}\left(\text{pool}\left(\mathbf{X}^{(H)}\right), \mathbf{g}\right), \qquad \hat{y} = \phi(\mathbf{g}') \tag{2}$$

where $\hat{y}$ denotes the predicted target property and $\phi$ represents the readout neural network. We note that for node-level target properties the pooling operator is omitted (eqn. 2).

Our EGNN encoder uses separate message-passing layers for processing invariant scalar features (e.g., atomic property descriptors) and 3D coordinate information based on the architecture first introduced by Satorras and coworkers[32]. The EGNN architecture uses no edge-based features beyond pairwise distances. The use of interatomic distances enables E(3)-equivariance at modest computational expense compared to architectures utilizing spherical harmonics and tensorial equivariant features.[32,96] The pairwise distances are defined by atomic coordinates, $\mathbf{z}_i$, and projected into a continuous space, $\mathbf{r}_{ij}$, via Gaussian basis functions along with optional multi-head attention, $G$ (eqn. 3):

$$\mathbf{r}_{ij} = G\left(\left\|\mathbf{z}_i^{(h)} - \mathbf{z}_j^{(h)}\right\|^2\right) \tag{3}$$

where the superscript $(h)$ denotes the hidden layer. Our architecture next updates node (eqn. 4) and edge (eqn. 5) embeddings through a series of message-passing operations, referred to as "sublayers", to update node and edge embeddings:

$$\mathbf{e}_{ij}^{(s)} = \phi_e\left(\mathbf{x}_i^{(s)}, \mathbf{x}_j^{(s)}, \mathbf{r}_{ij}\right) \tag{4}$$

$$\mathbf{x}_i^{(s+1)} = \mathbf{x}_i^{(s)} + \phi_n\left(\mathbf{x}_i^{(s)}, \mathrm{pool}_{j\in\mathcal{N}(i)}\left(\mathbf{e}_{ij}^{(s)}\right)\right) \tag{5}$$

where $\mathbf{e}_{ij}^{(s)}$ is the edge embedding between nodes $i$ and $j$ at sublayer $s$, $\phi_e$ and $\phi_n$ are small MLP networks dedicated to updating edge and node embeddings, and $\mathbf{x}_i^{(s)}$ is the node embedding of node $i$ at sublayer $s$. We emphasize that $\mathbf{x}_i^{(s)}$ (eqn. 5) does not include 3D atomic coordinates, which are stored separately as $\mathbf{z}_i$ (eqn. 6). After repeating for all $S$ sublayers, the latent embedding of coordinates is updated:

$$\mathbf{z}_i^{(h+1)} = \mathbf{z}_i^{(h)} + \mathrm{pool}_{j\in\mathcal{N}(i)}\left(\frac{\mathbf{z}_i^{(h)} - \mathbf{z}_j^{(h)}}{\sqrt{\mathbf{r}_{ij}} + 1}, \phi_z\left(\mathbf{x}_i^{(S)}, \mathbf{x}_j^{(S)}, \mathbf{r}_{ij}\right)\right) \tag{6}$$

where $\phi_z$ is a dedicated coordinate-update network. This entire process is then repeated for $H$ sets of layers (i.e., equivariant blocks), resulting in a final set of learned node embeddings $\mathbf{X}^{(H)}$ and coordinate embeddings $\mathbf{Z}^{(H)}$. These latent embeddings are pooled, concatenated to graph-level descriptors, and projected through a readout network in a manner identical to the 2D GNN architecture (eqn. 2).

To map the learned representations output by a GNN encoder to target variables, ElemeNet supports an MLP readout module standard in graph-based learning tasks. For increased expressivity, transformer-based readout is also implemented via multi-head self-attention and feed-forward sublayers across the embeddings learned by the GNN encoder. This enables dynamic weighting of different components in the molecular graph. While this approach facilitates more rapid learning of complex interdependencies and long-range interactions, it does so at the expense of increased computational cost during training. The modularity of ElemeNet's encoder-readout architecture results in a vast hyperparameter space, necessitating efficient optimization over proposed model configurations.

For a given encoder-readout architecture (e.g., a GNN encoder with MLP readout), hyperparameter optimization is handled by Optuna[80], a Bayesian optimization framework for navigating the high-dimensional space of potential activation functions, convolution operators, pooling functions, dropouts, neurons, number of layers, and network shapes (Supporting Information Table S5). Five nonlinear activation functions are available for all layers (i.e., ReLU, TanH, Leaky ReLU, GeLU, and SiLU), offering a range of smoothness properties. For 2D GNN encoders, five convolution operators (i.e., GraphConv, GCNConv, GINEConv, NNConv, and GAT) are also supported, along with mean, sum, and max pooling where applicable (i.e., only for graph-level prediction tasks). A key additional parameter to consider is network shape. Existing literature on representation learning has highlighted the benefits of compressing or expanding model latent space during training[97-100]. In addition to the number of neurons and hidden layers in the encoder and readout modules, we separately consider encoder and readout shapes as tunable hyperparameters, which may be defined in terms of the number of neurons per hidden layer either being constant, increasing, decreasing, "hourglass", or "pyramid". For the latter two cases, hourglass refers to the case where the neuron count decreases and then increases and pyramid corresponds to the case where the neuron count increases then decreases (Figure 2). These hidden layer shapes enable exploration of network structure and information bottlenecks within the encoder and readout modules, a design choice often overlooked in existing molecular machine learning codes. Bayesian optimization over this vast design space is central to ElemeNet's capability to efficiently discover high-performing architectures for diverse prediction tasks in an automated fashion. Together with the modular encoder-readout design and robust initial representations, ElemeNet is well-suited as a general-purpose framework for molecular machine learning.

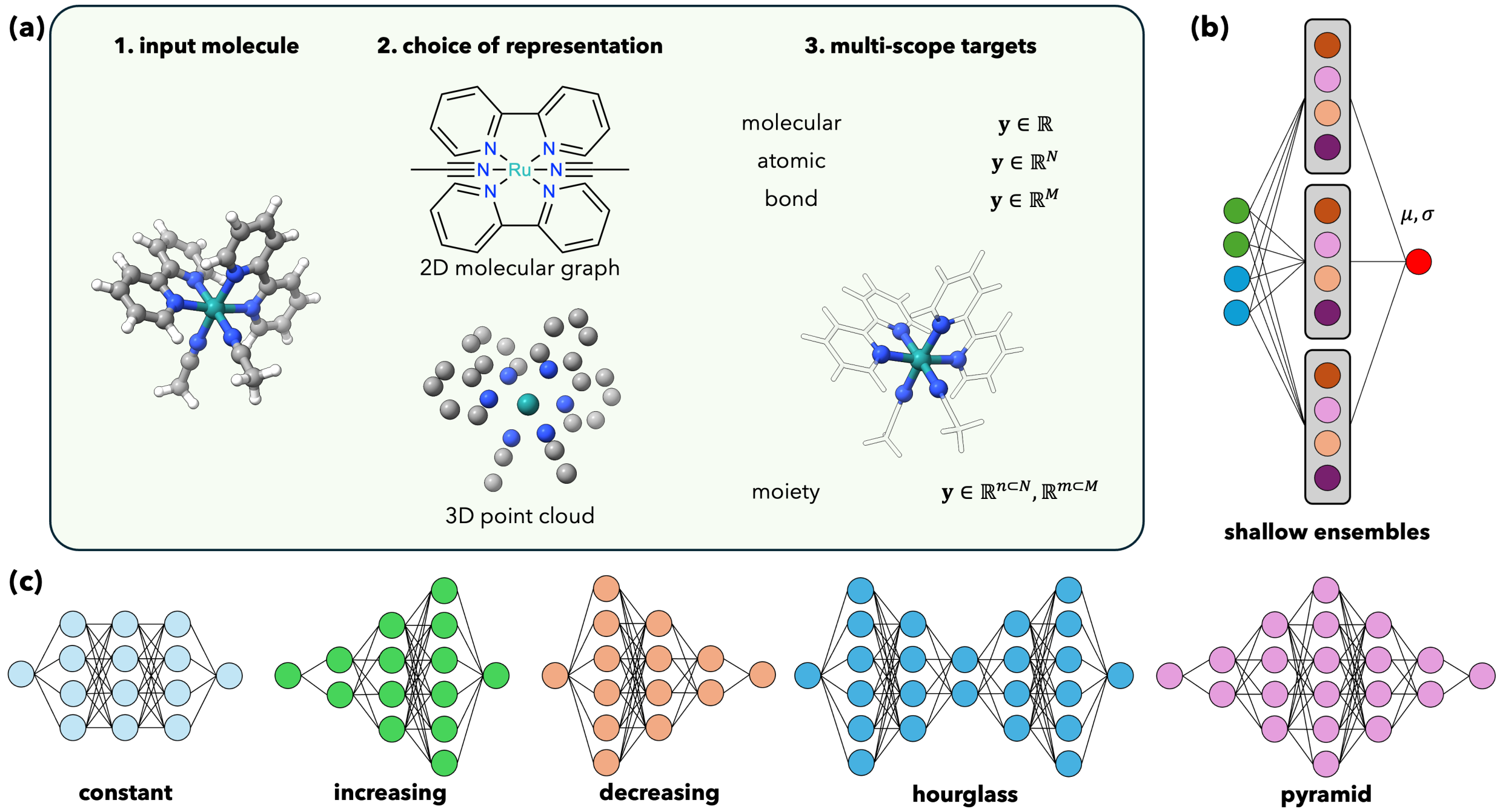

**Figure 2.** (a) Summary of different representations and targets available in ElemeNet. A given molecular input may be transformed to 2D or 3D molecular graphs which are used to predict targets at the molecular (graph), atomic (node), bond (edge), or moiety (subgraph) level. (b) Shallow ensemble readout layers supported in ElemeNet. Given a shared model trunk, the final readout layer is replaced with a last-layer ensemble that is independently trained. The mean ($\mu$) and standard deviation ($\sigma$) in predictions across the ensemble heads are used to parameterize a predictive distribution. The resulting negative log likelihood is used as the loss function to be minimized in training the overall model. (c) Supported network shapes in ElemeNet. In networks with "constant" shape, the number of neurons is the same for every hidden layer. "Increasing" shapes see the hidden size gradually get larger with each layer, while hidden layer size in "decreasing" networks steadily decreases. "Hourglass" and "pyramid" shapes are more sophisticated schemes in which the hidden size is sequentially expanded and contracted.

**3b. Multiscale Learning Tasks.**

ElemeNet is designed to accommodate diverse prediction tasks across different chemical domains and target properties. Specifically, our framework supports both regression and classification tasks. For regression tasks, a model predicts continuous, real-valued properties (e.g., electronic energy, band gap, solubility) by applying a loss function directly to final-layer model outputs. For classification, ElemeNet supports both binary and multiclass prediction problems. In binary classification (e.g., predicting whether a molecule will be reactive or inert), a sigmoid activation function transforms final-layer logits to probabilities, with class membership determined by thresholding predictions at 0.5. Similarly in multiclass classification, a softmax activation function transforms logits into a length-$K$ vector of per-class probabilities, with class membership assigned to the class with the highest predicted probability. All capabilities within ElemeNet, including featurization, model architectures, and hyperparameter tuning, are agnostic to task type and are equally supported for regression, binary classification, and multiclass problems. ElemeNet also supports multitask learning, in which a single shared encoder processes molecular features and generates learned representations, and multiple independent readout heads predict each target property from these shared embeddings. Multitask training is supported for both regression and binary classification problems. In both cases, the shared encoder and task-specific readout heads are trained jointly, with loss aggregated per-target for shared backpropagation. This approach exploits shared chemical structure and is particularly useful in the context of limited training data or correlated target properties[101-105] (e.g., polarizability and solubility[106], frontier orbital energies[48]).

Beyond task type (i.e., regression vs. classification), ElemeNet supports predictions across multiple scopes within a single molecular graph (Figure 2). Graph-level properties are the most common target in molecular machine learning, where a single target or set of targets is assigned to a single molecule (e.g., toxicity, total electronic energy). To obtain a molecule-level embedding from node-level representations, a pooling function is applied to aggregate contributions from each node into a fixed length vector for each molecule. The code supports node-level predictions corresponding to atom-centered targets (e.g., partial charges, condensed Fukui functions), which are defined similarly but with the absence of a pooling function used for graph-level targets. The permutation equivariance of graph neural networks without pooling results in a length-$N$ vector of node-level predictions matching the order of atoms specified in the input molecule.[107] Finally, ElemeNet supports training models on bond-level targets (e.g., bond lengths, natural bond order outputs) through dedicated edge readout architectures. As the learned embeddings generated by message-passing are node-centered, an additional projection step is required to map these learned representations to edge-centered targets. ElemeNet accomplishes this through last-layer projection, in which edge features are concatenated to learned node embeddings for each edge $(i, j)$ and passed through an MLP to generate bond-centered predictions. All existing model capabilities (e.g., transformer readout, shallow ensembles) are compatible with the edge-level architecture.

While atom, bond, and molecule-level targets are ubiquitous in molecular machine learning, many properties of interest are localized to a specific chemical motif (i.e., "moiety"). Such targets exist in between the scope of single atoms and entire molecules. To accommodate this unique requirement, ElemeNet introduces moiety-level predictions, defined as a local subgraph centered on a central node index and extending to all atoms and bonds within a $k$-hop neighborhood (Figure 2). Node representations are updated by message-passing using the entire molecule as context, with loss localized to the target moiety by masking all atoms beyond the $k$-hops subgraph. Within a defined chemical moiety, the target property may still be defined on a node, edge, or (sub)graph basis, where the set of nodes and edges is confined to those within the *k*-hops neighborhood. We note the distinction between this masked-subgraph-level inference and global graph-level inference, which places equal weight on all atomic neighborhoods rather than confining the loss to the chemically meaningful motif. Moiety-level predictions are particularly beneficial for chemistry, where target properties are often centered upon an atom, functional group, or active site of interest. Representative use cases include metal–ligand bonding interactions in transition metal complexes, physiochemical properties of specific functional groups (e.g., conformational strain, p$K_a$), and active site reactivity in enzyme models. Combined with support for multitask and multiscale learning, we anticipate the introduction of moiety-level targets will make ElemeNet particularly useful in molecular machine learning.

**3c. Uncertainty Quantification.**

Uncertainty quantification (UQ) is a key component to machine learning workflows when it provides valuable insights regarding when model predictions are supported by training data. Bayesian approaches placing posterior distributions over model parameters are limited by their steep computational cost, while distribution-free uncertainty estimates are prone to being poorly calibrated with prediction error.[71] ElemeNet provides uncertainty estimates in two tiers. First, low-cost heuristic uncertainty measures are computed from a single forward pass during model training. Second, we implement a scalable ensemble-based approach that produces predictive uncertainties with minimal computational overhead (Figure 2). The first of the heuristic methods is latent space distance, a widely used uncertainty estimate popular for its simplicity of implementation and minimal computational cost.[64-66,108] For a given molecule test-set molecule $x^*$, the distance between its latent representation and the closest point $x$ in the training set latent space is the latent space distance. ElemeNet reports latent space distances in both Euclidean and cosine space, taken as the distance to the $N$ nearest training-set points in the learned embedding. Euclidean and cosine forms are both reported because embedding magnitude or direction may carry more information about novelty, depending on the encoder's regularization. While latent space distances have an intuitive interpretation (i.e., that a model is expected to perform best on test data "near" the training data), they are not considered a calibrated uncertainty metric, as they do not quantify expected error magnitude. They also do not impact the model training process, being calculated only during inference and going unused during backpropagation. It has been shown that latent distances can be partially calibrated against predictive error through engineering of the embedding metric or through post-hoc procedures.[67] However, these calibration steps are not part of the standard distance computation and require additional analysis applied to held-out data. As such, latent space distances are entirely reliant on the quality of the learned representation.[67-69] For classification tasks, additional uncertainty metrics are available in the form of information entropy (i.e., the Shannon entropy of the predicted class distribution, eqn. 7).[109] For a $K$-class classifier with predicted probabilities $p_k(x)$, entropy is defined as:

$$H = -\sum_{k=1}^{K} p_k \log p_k \tag{7}$$

Information entropy captures the notion that diffuse probability mass (i.e., predictions far from 0 or 1) should be treated as more uncertain, and in practice serves as a useful heuristic for flagging low-confidence predictions. However, information entropy does not directly reflect predicted error magnitude – it is a property of the predicted distribution, not a calibrated measure of how often the model is right or wrong[110-112]. Together, latent space distance and information entropy serve as rapid and informative diagnostics for uncertainty, they are heuristic measures of embedding-based novelty not tied to the training objective. For uncertainty quantification more formally grounded in predictive performance, we turn to ensemble-based approaches.

Ensembles are widely recognized as among the most mathematically rigorous and empirically successful strategies for uncertainty quantification in deep learning. Fully Bayesian strategies that define posterior distributions over model parameters exhibit excellent calibration but are often infeasible due to the computational overhead of training multiple fully independent models.[72-75,77,113] To achieve well-calibrated uncertainty estimates while avoiding the limitations of so-called deep ensembles, ElemeNet utilizes the recently introduced approach of using shallow ensembles to quantify prediction uncertainty.[76] In shallow ensembles, all network weights up to the penultimate layer are shared, while the final prediction layer is replicated $N$ times (eqn. 8). The result is a model producing $N$ predictions for each input in a single forward pass:

$$\hat{\mathbf{y}} = \{\hat{y}_1, \hat{y}_2, \dots, \hat{y}_N\} \tag{8}$$

where $\hat{y}_n$ is the prediction generated by ensemble member $n$. Because only the final layer is replicated, computational overhead is small, while the diversity across heads provides a practical approximation to epistemic uncertainty. For a detailed derivation and benchmarking studies, the reader is referred to Kellner and Ceriotti.[76]

For regression tasks, ensemble predictions are interpreted as samples from a predictive distribution. We parameterize a Gaussian centered around the ensemble mean, $\mu$, and assign a variance, $\sigma^2$, based on the spread across the $N$ prediction heads:

$$\mu = \frac{1}{N}\sum_{n=1}^{N} \hat{y}_n, \qquad \sigma^2(x) = \frac{1}{N-1}\sum_{n=1}^{N} (\hat{y}_n - \mu)^2 + \epsilon \tag{9}$$

with a small offset, $\epsilon$, included for numerical stability (eqn. 9). The model is trained by minimizing the negative log-likelihood (NLL) of the observed regression targets under this Gaussian:

$$\mathcal{L}_{\text{NLL}} = \frac{1}{2}\left[\log\sigma^2 + \frac{(y-\mu)^2}{\sigma^2} + \log 2\pi\right] \tag{10}$$

Dropping the constant $\frac{1}{2}\log 2\pi$ term:

$$\mathcal{L}_{\text{NLL}} = \frac{1}{2}\left[\log\sigma^2 + \frac{(y-\mu)^2}{\sigma^2}\right] \tag{11}$$

where $\mathcal{L}_{\text{NLL}}$ denotes the NLL loss. This formulation has an intuitive interpretation: the model is penalized both for being wrong and for having poorly calibrated uncertainties (i.e., predicting large $\sigma^2$ everywhere increases $\log\sigma^2$, while predicting small $\sigma^2$ everywhere is penalized by the $1/\sigma^2$ prefactor, eqn. 10 and 11). Unlike distance-based heuristics, the NLL objective directly ties the learned uncertainty to predictive performance during training, encouraging uncertainties that are meaningful with respect to observed errors. While latent space distance as a measure of distinctness of a test point to the embedding space can be effective at identifying out-of-distribution inputs, it exhibits no formal relationship to error magnitude and depends entirely on the structure of the learned embedding space. In contrast, shallow-ensemble NLL regression yields a predictive distribution with a variance parameter optimized to reflect residual error.

We compare the performance of ElemeNet regression models trained with and without shallow ensembles on both the QM9[47] and tmQMg[84] datasets and observe strong improvements in both error calibration and predictive accuracy (Figure 3). 2D GNN models with MLP and transformer readout and identical hyperparameters and data splits are trained without ensembles and with ensemble sizes of 2, 4, 8, 16, 32, and 64 to predict QM9 atomization energy (Supporting Information Table S6 and S7). The same procedure is applied with EGNN models trained on tmQMg to predict the magnitude of the dipole (Supporting Information Table S8 and S9). We compare the resulting models in terms of predictive performance and uncertainty calibration, defined as the Spearman rank correlation between an uncertainty metric and test-set mean absolute error (MAE). Spearman rank correlation is the standard metric in UQ literature because it makes no assumptions about whether the relationship between error and uncertainty is linear (i.e., homoscedastic) or nonlinear (i.e., heteroskedastic). In both cases, we observe models trained with ensembles to exhibit error calibration exceeding latent space distance baselines (Figure 3). Prior work has observed increased performance of latent space distances when averaging among $N > 1$ nearest neighbors.[64,65] As such, we calculate latent space calibration among $N$ nearest training-set neighbors, considering previously recommend values of $N = 10$ and $N = 200$ in addition to the default value of $N = 1$. In line with previous work, we observe that increasing $N$ generally yields increased calibration of latent space distance, though the results vary among dataset, architecture, and target property (Supporting Information Table S6–S9).

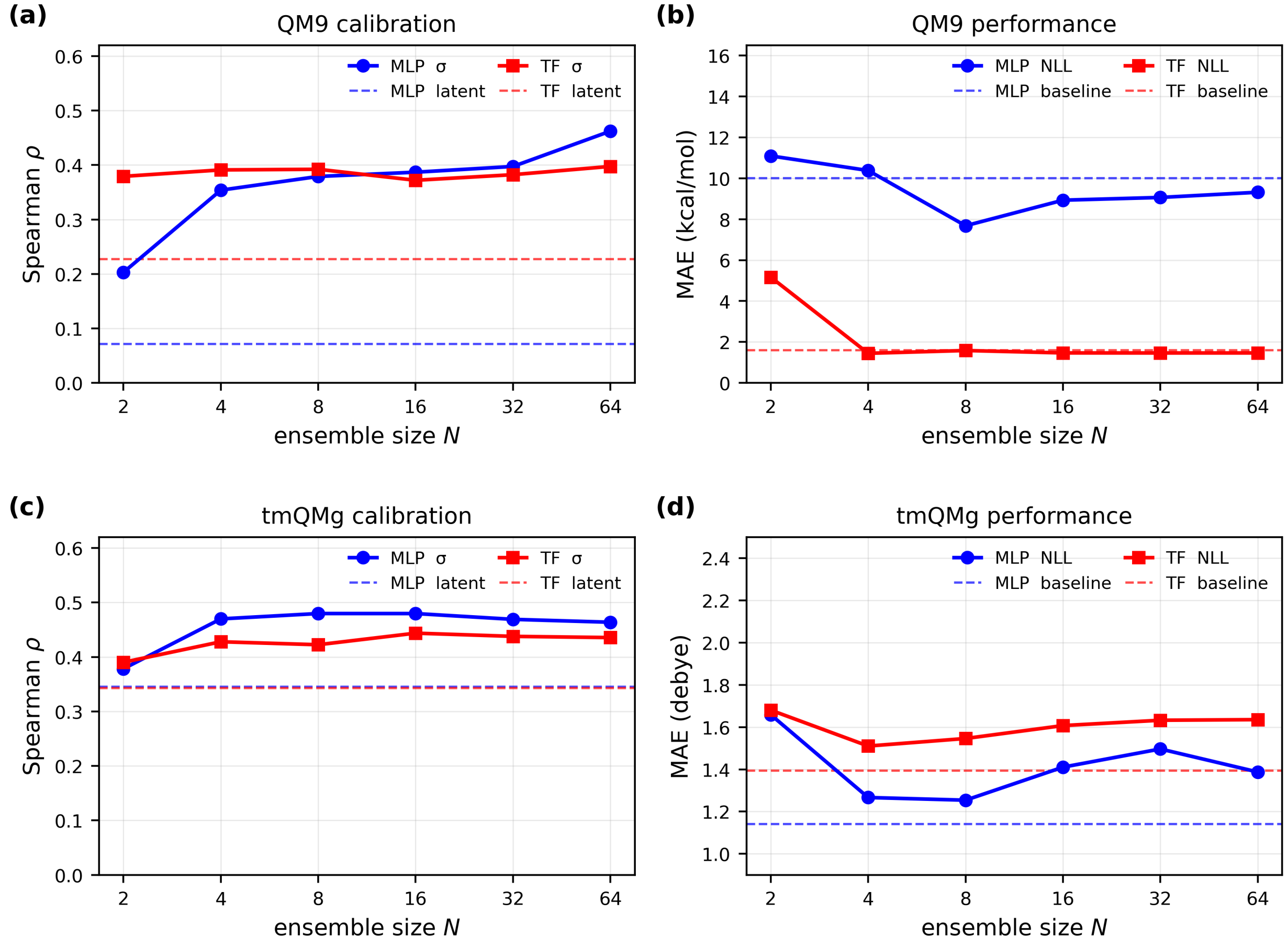

**Figure 3.** Comparison of calibration and predictive performance in regression models trained with and without shallow ensembles. Results for multilayer perceptron (MLP) and transformer (TF) readout layers are shown separately. (a) Spearman rank correlation ($\rho$) between ensemble standard deviation ($\sigma$) and mean absolute error (MAE) on QM9 atomization energy across ensemble sizes. Comparison is made to latent space distance baselines, with higher values indicating better calibration. (b) MAE of 2D GNN models trained on QM9 atomization energy compared against ensemble-free baselines. (c) Correlation between ensemble standard deviation and MAE on tmQMg dipole magnitude across ensemble sizes. Comparison is made to latent space distance baselines, with higher values indicating better calibration. (d) MAE of EGNN models trained on tmQMg dipole magnitude compared against ensemble-free baselines.

A noteworthy finding is that increasing ensemble size also increases the calibration of latent space distance, suggesting that ensembling improves the calibration of multiple uncertainty metrics, though ensemble variance remains a more highly calibrated metric (Supporting Information Table S6–S9). Models trained with NLL loss also exhibit MAE comparable to those of ensemble-free methods, indicating this increase in uncertainty calibration does not compromise predictive performance (Figure 3, Supporting Information Table S6–S9). We note slightly increased errors in dipole for the tmQMg models, which may be attributable to the lack of hyperparameter tuning in any of these tests; these tests are deferred to further benchmarking studies (Section 3e). In practice, we observed ensemble-based uncertainties to exhibit stronger correlation with error magnitude and result in better-performing models at a small increase to computational cost on most learning tasks (Supporting Information Tables S6–S9).

The original work introducing direct propagation of shallow ensembles for uncertainty quantification was limited to regression tasks. In this work, we additionally explored extending the shallow ensemble framework to classification tasks. The most natural analogue of NLL regression to binary classification is to parameterize a Beta distribution from the mean and variance of ensemble predicted probabilities and minimize the negative log-likelihood during training. However, the resulting Beta-Bernoulli NLL can be shown to reduce exactly to binary cross-entropy (Supporting Information Text S4). Similarly for multiclass classification, parameterization of a Dirichlet distribution results in a Dirichlet-categorical loss equivalent to standard cross-entropy (Supporting Information Text S5). Unlike in the regression case, no variance term remains as each loss function collapses to cross-entropy, and the ensemble reduces to a mean prediction across heads (Supporting Information Text S4 and S5). UQ for classification tasks is better suited to a mean-of-probabilities (MoP) ensembling scheme,[114,115] in which each head produces a logit (binary) or logit vector (multiclass), which is passed through the task-specific readout activation function (i.e., sigmoid or softmax, eqn. 12 and 13) and averaged across heads:

$$\hat{p} = \frac{1}{N}\sum_{n=1}^{N} \text{sigmoid}(\hat{l}_n), \qquad \text{binary} \tag{12}$$

$$\hat{\mathbf{p}} = \frac{1}{N}\sum_{n=1}^{N} \text{softmax}(\hat{\mathbf{l}}_n), \qquad \text{multiclass} \tag{13}$$

where $\hat{p}$ and $\hat{\mathbf{p}}$ are predicted binary and per-class probabilities, $\hat{l}$ and $\hat{\mathbf{l}}$ are predicted binary and per-class logits, and $N$ is the number of ensemble heads. The model is then trained on standard binary or categorical cross-entropy applied to this mean prediction. This formulation has two important properties. First, because aggregation occurs in probability space rather than logit space, every ensemble head receives independent gradient signal through the chain rule, avoiding the gradient collapse of NLL-style ensemble classifiers. Second, by Jensen's inequality, mean-of-probabilities is in general not equal to a sigmoid- or softmax-of-mean and is instead pulled toward the uniform distribution when heads disagree.[116,117] MoP therefore softens overconfident predictions where the ensemble disagrees while preserving sharp predictions where it agrees. As a per-prediction uncertainty estimate, the probability-space ensemble standard deviation $\sigma_p$ is reported alongside the mean prediction.

We next tested whether MoP offers any measurable benefit to models trained with standard cross-entropy loss. We repeated our above analysis by training a series of 2D GNN models with MLP and transformer readouts on the classification tasks of coordinating atoms (binary) and coordination number (multitask) prediction on the pydentate dataset, sweeping across the ensemble sizes ranging from 2 to 64 (Figure 4). We use splits derived from previous work[85] (i.e., dropping the 0.1% of ligands which were unparsable, Supporting Information Table

S10), omitting all ligands with phosphorous coordinating atoms to curate a held-out set of out-of-distribution (OOD) molecules (Supporting Information Table S11 and S12).

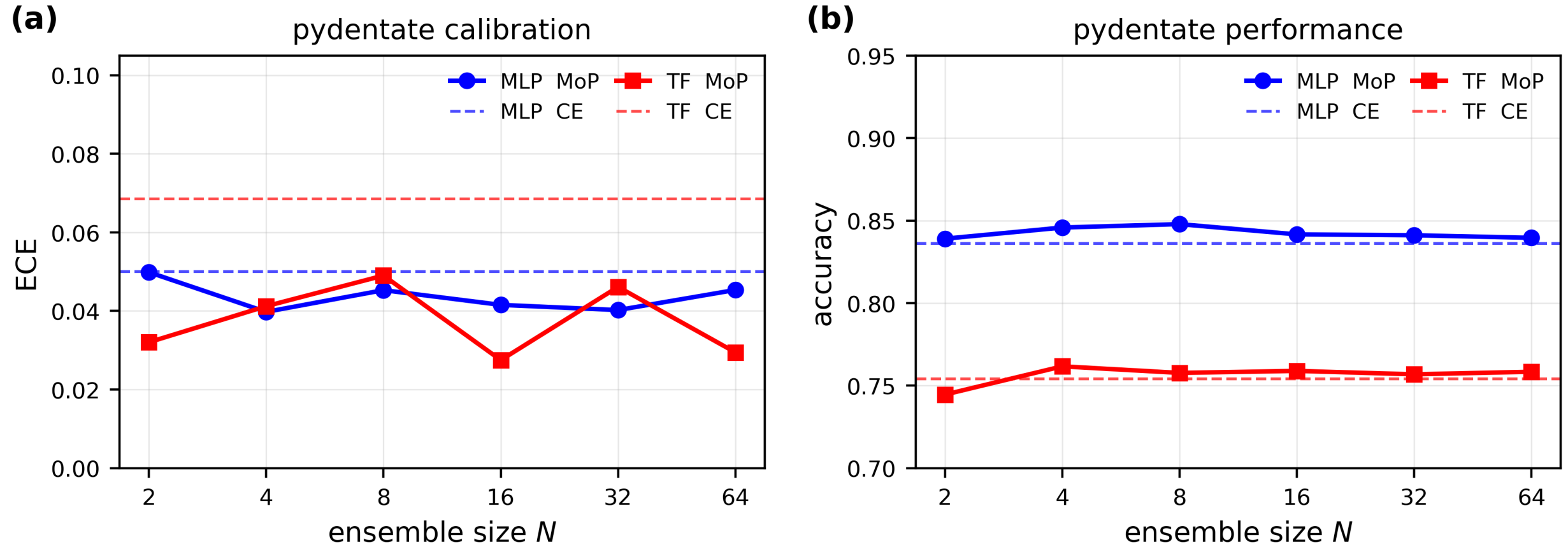

**Figure 4.** Comparison of calibration and predictive performance of classification models trained with and without shallow ensembles. Results for multilayer perceptron (MLP) and transformer (TF) readout layers are shown separately. (a) Expected calibration error (ECE) between confidence and accuracy of predicted probabilities on pydentate coordination number across various ensemble sizes. Comparison is made to cross-entropy baselines, with lower values indicating better calibration. (b) accuracy of 2D GNN models trained on pydentate coordination number compared against cross-entropy baselines.

Across binary and multiclass classification benchmarks, MoP matches standard cross-entropy baselines on predictive accuracy and area under the receiver operating characteristic (AUROC) at ensemble sizes from $N = 2$ to $N = 64$ (Supporting Information Table S11 and S12). This confirms that the additional last-layer width does not destabilize training. Meanwhile, uncertainty calibration under MoP improves substantially on both in-distribution and out-of-distribution (OOD) data (Figure 4). We judge ensemble performance through expected calibration error (ECE), which is a standard metric for quantifying the miscalibration of deep learning models. For multiclass classification with transformer readouts with 64 heads, MoP reduces ECE by 57% relative to standard cross-entropy on the in-distribution test set and 34% on set-aside OOD data (Supporting Information Table S12). Nevertheless, this observation is sensitive to the number of heads and is not monotonic (Supporting Information Table S12). Because MoP is rank-preserving in expectation, AUROC-based metrics for misclassification detection and OOD identification are unchanged relative to standard cross-entropy. The improvement is in probability calibration, useful for learning tasks when predicted probabilities will be subjected to threshold-based decision making, such as in variable-threshold evaluation metrics[57-61], active learning[62], and reinforcement learning[63]. We therefore recommend the MoP metric introduced here as an effective shallow-ensemble analogue to NLL-regression.

**3d. Software Overview.**

To maximize usability among diverse chemical elements and domains, ElemeNet is compatible with most common molecular structure formats. 2D molecular graphs provided in cif, cml, mol, mol2, sdf, SMILES, and xyz formats are all supported. Similarly, 3D molecular graph input formats are also supported for all of these file formats, with the exception of SMILES strings, which by definition encode only 2D connectivity. In the case of xyz files, which represent 3D coordinates without bonding information, connectivity is inferred from pairwise distances between atoms and a predefined dictionary of bonding thresholds (Supporting Information Text S1). The `mol_to_graph` function handles the conversion of molecular graph strings into ElemeNet graph objects. ElemeNet supports the use of either explicit or implicit hydrogen atoms, and optional node and edge features extracted by RDKit are also available (Supporting Information Table S13).

Inspired by recent work on resonance-invariant graph representations[118], bond order information may be optionally dropped. After constructing a graph object, node, edge, and graph-level attributes are added (Supporting Information Table S14). Different functions are responsible for defining the initial node, edge, and graph representations for a given molecule, and support the specification of additional user-provided features at the atom (e.g., partial charges), bond (e.g., Wiberg bond indices), or molecule (e.g., solubility) level. Additionally, molecular charge and spin multiplicity may be specified as user-provided inputs at the graph level (Supporting Information Table S14). In the absence of user-provided charge and spin information, these variables are inferred from the number of electrons by assuming neutral atoms and a low-spin state for the overall molecule.

The code reads all molecular graph inputs and target properties from a .csv file, where the columns containing the molecular graphs, target properties, and user-provided features are specified as inputs (Supporting Information Table S14). ElemeNet supports the specification of predefined training, validation, and test sets provided as separate .csv files. A single .csv containing the full dataset and a specified training, validation, and test split is also supported. In the case of imbalanced datasets for classification tasks, ElemeNet supports the optional preservation of the distribution of class labels in the full dataset when splitting the data into training, validation, and test sets. Additionally, the user may specify the data split (e.g., all in train, or all in test). This is particularly useful in the case of high similarity among data points, where the inclusion of similar molecules in training and test splits may result in data leakage and artificially inflated test set performance.

Model construction relies on an encoder and readout routine where the graph neural networks and multilayer perceptron or transformer architectures are defined, respectively (Supporting Information Table S14). Once the encoder and readout configurations are defined, the combined network is built, defining masks as necessary to subselect nodes, edges, and subgraphs for atom, bond, and moiety-level predictions (Figure 5). The constructed model is then trained by a routine that accepts the model, optimizer, and loss configurations as inputs (Figure 5). Supported optimizers include stochastic gradient descent (SGD) and adaptive moment estimation with decoupled weight decay (AdamW), among others (Supporting Information Table S14). Numerous loss functions are also defined, and include mean squared error, mean absolute error, binary cross-entropy, and cross-entropy. Ensemble-based loss functions are additionally defined for regression, binary classification, and multiclass classification tasks (Supporting Information Table S14). Bayesian hyperparameter optimization is carried out over a space of encoder and readout hyperparameters by default (Figure 5). For a cursory, less rigorous hyperparameter search, an available flag restricts the space of hyperparameters over which optimization is performed (Supporting Information Table S15).

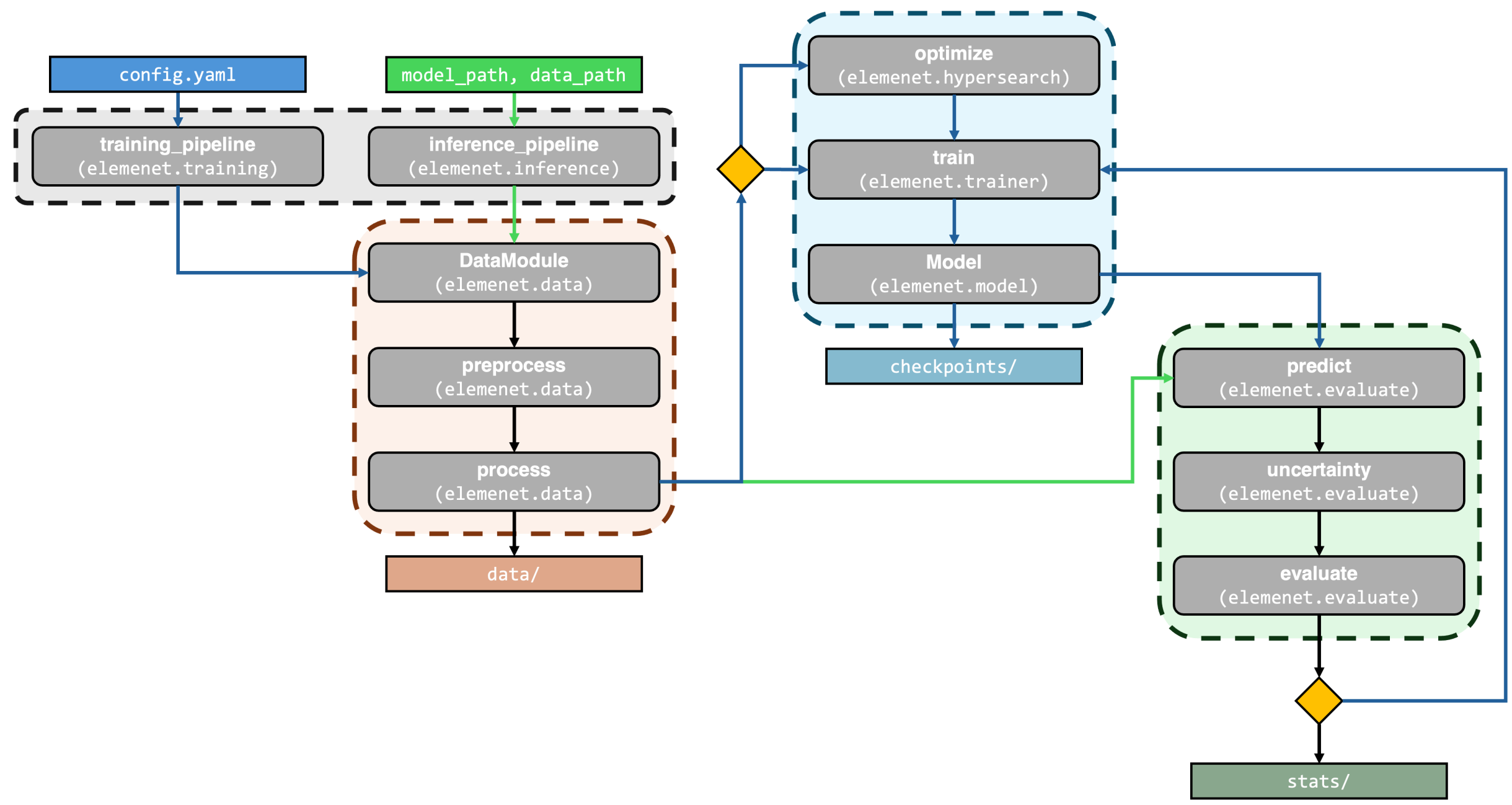

**Figure 5.** Block flowchart structure of the ElemeNet codebase. The four subsections with dotted outlines correspond to wrapper functions (gray), preprocessing (orange), model training (blue), and inference (green). Within each of the four subsections, key functions and classes are shown in gray with their corresponding module files in parentheses. For example, the `DataModule` class for creating graph objects is defined in the `elemenet.data` module. Inputs and logical flow for `training_pipeline` and `inference_pipeline` are colored separately in blue and green, respectively. Operations shared between the two pipelines are indicated with black arrows. Hyperparameter optimization is optional and gated by a yellow diamond. When active, the `optimize` function is called to iteratively train and evaluate models with different hyperparameters. When inactive, `optimize` is bypassed and a single production model is trained. Outputs to preprocessing, model training, and inference are stored separately in the `data/`, `checkpoints/`, and `stats/` directories.

ElemeNet offers two wrapper functions for end-to-end training and inference pipelines. The first is `training_pipeline`, which wraps all preprocessing, hyperparameter optimization, and model training steps and keywords in a single function (Figure 5). The second wrapper is `inference_pipeline`, which preprocesses a given dataset and runs inference using a trained model, returning model predictions, evaluation metrics (e.g., mean-squared error, accuracy), and uncertainties (e.g., standard deviation, latent space distances, information entropy) associated with each prediction (Supporting Information Tables S16 and S17). Both are available through the command-line interface. The command `elemenet_train` requires only a .yaml file specifying the model configuration (Supporting Information Table S14). The command `elemenet_inference` accepts paths specifying the trained model and where to save inference results along with the target property, the molecular graph label, and graph format. Fine-tuning of a pretrained model is also easily accessible from the command line via a resume feature. ElemeNet relies on several key external packages for preprocessing, training, and hyperparameter optimization. Molecular structure inputs are generally parsed within ElemeNet, with the exception of SMILES strings which are read using RDKit[81] and cif files in which the periodic coordinates are handled by molSimplify[119]. ElemeNet is built on PyTorch Geometric[79], while Bayesian optimization over a defined hyperparameter space is performed with Optuna[80]. ElemeNet also supports distributed training across multiple GPUs for accelerated hyperparameter tuning and model training.

**3e. Benchmarking Studies.**

To evaluate the performance of ElemeNet on property prediction tasks across diverse chemical domains, we select four representative datasets from organic, inorganic, coordination, and biological chemistry encompassing graph-, node-, edge-, and subgraph-level prediction targets. Our objective is to show that models competitive with the state of the art can be trained across each of these settings using a single unified codebase with minimal user effort. More precise fine-tuning beyond that carried out here would likely be necessary to obtain state-of-the-art performance on an individual dataset. Accordingly, all hyperparameter searches were limited (i.e., only 25 trials of 100 epochs each), and we did not separately tune hyperparameters for every architecture-dataset combination. More rigorous searches with larger batch sizes and longer training would yield further improvements. Production models trained with ElemeNet should deploy such practices.

We first consider QM9, a widely used DFT dataset of 133,885 organic molecules containing up to nine heavy atoms (C, N, O, F).[47] We separately train models to predict atomization energy, HOMO-LUMO gap, dipole magnitude, and polarizability. We vary both the encoder (2D or E(3)-equivariant GNN) and readout (MLP or transformer) architecture. We additionally consider all model configurations with and without shallow ensembles, resulting in 32 total baselines reported (i.e., four architectures, two loss functions, four target properties). For each combination, hyperparameters are selected for each model against validation loss, and the identified optimal configuration is retrained with early stopping (see Methods). To benchmark against an established machine learning workflow for organic chemistry, we compare to Chemprop, and we use the same data splits reported by Heid et al.[51,52] For each 2D GNN considered, we compare the performance of models trained with and without optional SMILES features, reporting the results of the best-performing model with predictions averaged across a deep ensemble of 5 models manually prepared and trained using five different random seeds, emulating the procedure performed in the reported Chemprop baselines (see Methods, Supporting Information Table S13).[51,52]

The full comparison of model architectures across QM9 reveals two consistent trends. First, the EGNN encoder substantially outperforms the 2D GNN on all target properties, with the effect being most pronounced for target properties where 3D geometry is expected to be important. The best-performing EGNN outperforms the corresponding most accurate 2D GNN by a factor of only 1.6 for HOMO-LUMO gap, while MAEs are reduced by over sixfold for the dipole by the EGNN (Table 1). This trend is consistent with chemical intuition, as dipole is highly dependent on three-dimensional structure that a 2D topological representation cannot fully encode. Second, shallow ensembling is generally observed to result in slight increases in MAE for some properties (e.g., atomization energy), reinforcing tradeoffs documented in the literature between test-set accuracy and calibrated uncertainty (Table 1). Nevertheless, for other properties, ensemble mean errors are lower than those obtained without ensembling (Table 1). We note that the hyperparameters of all models were optimized separately but with equal search spaces and maximum allowed trials, making direct comparisons difficult. Evaluating the effect of ensemble size under a fixed model architecture revealed consistent benefits to training under the shallow ensemble paradigm (Figures 3 and 4). As ensemble-based methods exhibit increased variance, an increased optimization budget is recommended for production models trained with shallow ensembles.

**Table 1.** QM9 property prediction performance for all combinations of encoder (2D GNN, EGNN), readout (MLP, TF), and shallow ensemble size (1, 16). All values are test-set MAEs on the splits of Heid et al.[51] Hyperparameters were optimized for each combination over 25 trials of 100 epochs each. Each model was trained with the optimal configuration for up to 1,000 epochs with early stopping. Chemprop model performance is from Heid and Graff et al.[51,52] Chemprop dipole and polarizability were reported using a multitask model. The best-performing ElemeNet model for each task is bolded.

| **encoder** | **readout** | **ensemble size** | **atomization energy (kcal/mol)** | **HOMO-LUMO gap (hartree)** | **dipole (debye)** | **polarizability ($bohr^3$)** |
|---|---|---|---|---|---|---|
| 2D GNN | MLP | 1 | 1.15 | $4.05\times10^{-3}$ | 0.441 | 0.322 |
| 2D GNN | MLP | 16 | 1.35 | $4.19\times10^{-3}$ | 0.401 | 0.271 |
| 2D GNN | TF | 1 | 1.08 | $3.79\times10^{-3}$ | 0.373 | 0.248 |
| 2D GNN | TF | 16 | 1.16 | $4.20\times10^{-3}$ | 0.431 | 0.243 |
| EGNN | MLP | 1 | **0.39** | $\mathbf{2.17\times10^{-3}}$ | 0.109 | **0.102** |
| EGNN | MLP | 16 | 0.55 | $2.82\times10^{-3}$ | 0.092 | 0.104 |
| EGNN | TF | 1 | 0.84 | $2.51\times10^{-3}$ | **0.060** | 0.197 |
| EGNN | TF | 16 | 0.55 | $2.81\times10^{-3}$ | 0.077 | 0.116 |
| Chemprop v1 (2D) | | | 1.11 | $3.14\times10^{-3}$ | 0.326 | 0.231 |
| Chemprop v2 (2D) | | | 1.02 | $3.12\times10^{-3}$ | 0.339 | 0.227 |

Selecting the best model per target, we find that ElemeNet is competitive with both the original (v1) and most recent (v2) releases of Chemprop. Our 2D GNNs are competitive with the corresponding Chemprop models on all properties, while our model trained with transformer readout outperforms Chemprop v1 on atomization energy, highlighting the benefits of attention mechanisms in molecular property prediction (Figure 6, Table 1). Our 3D models outperform all Chemprop baselines across all four target properties considered (Table 1). We note that the Chemprop dipole and polarizability values are only reported using a single multitask model[51,52], whereas the ElemeNet results are obtained from single-task training.

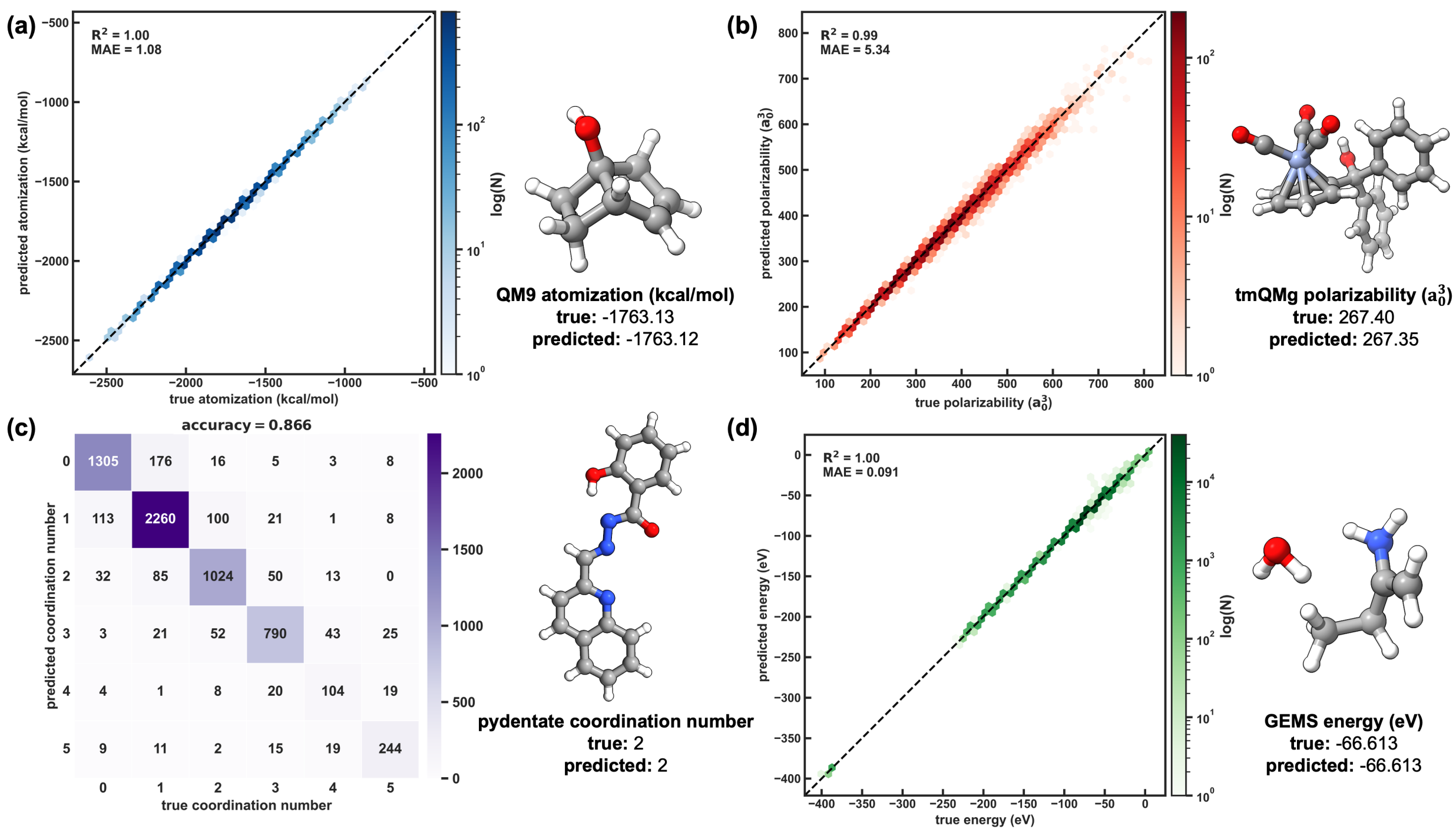

**Figure 6.** Representative test-set performance of ElemeNet models trained across diverse chemical domains and property prediction tasks. (a) Parity plot of a model (2D GNN encoder, transformer readout) trained on QM9 to predict atomization energy from SMILES inputs. (b) Parity plot of a model (EGNN encoder, transformer readout) trained on tmQMg to predict polarizability from xyz inputs. (c) Confusion matrix of a model (2D encoder, MLP readout, 16 last-layer ensembles) trained on the pydentate dataset to predict coordination number from SMILES inputs. (d) Performance of a model (EGNN encoder, transformer readout, 16 last-layer ensembles) trained on GEMS to predict electronic energy from xyz inputs. Insets depict representative molecules and corresponding true and predicted properties. Atoms are colored as follows: H in white, C in gray, N in blue, O in red, Cr in light blue.

We next consider tmQMg[84], a dataset of transition metal complexes extracted from the Cambridge Structural Database[83] and optimized with DFT, to evaluate ElemeNet on inorganic chemistry where 3D structure and metal coordination heavily influence properties. We evaluate performance on HOMO-LUMO gap, dipole, and polarizability, again comparing the same combinations of encoders, readouts, and ensembles. To alleviate the computational burden of exhaustive hyperparameter optimization and investigate the transferability of hyperparameters across model architectures, we tune hyperparameters only for the 2D GNN. We then transfer the optimal values to the EGNN with MLP readout, setting EGNN-specific parameters absent from the 2D search space to their defaults (see Methods). For the EGNN with transformer readout, we simply utilize the default hyperparameters in ElemeNet to evaluate performance under system defaults. To illustrate ElemeNet's multitask capability, we additionally train a single EGNN model with transformer readout on all three targets simultaneously (denoted EGNN*). This configuration is intended as a capability demonstration rather than a competitive single-task baseline.

Similar to the QM9 tests, the EGNN encoder outperforms the 2D GNN on every target, reinforcing the value of 3D-aware representations for transition metal chemistry. Ensembling provides marginal and inconsistent gains at this dataset size and hyperparameter search depth. The multitask EGNN* achieves accuracy within a factor of two of the dedicated single-task EGNN models on each target, despite using default hyperparameters and no architecture-specific tuning (Table 2). We next compared the best ElemeNet 2D and 3D models to the performance of the 3D architectures benchmarked by Kneiding et al., which span a range of representations including line-graph models and spherical harmonic equivariant networks.[84] We observed the best-performing ElemeNet outperforms the average tmQMg model on HOMO-LUMO gap ($8.37\times10^{-3}$ vs. 0.0104 hartree MAE), the dipole (0.97 vs. 1.10 debye MAE) and polarizability (5.34 vs. 5.62 bohr$^3$ MAE) target properties (Table 2, Figure 6, Supporting Information Table S18).[84] Given that no architecture- or dataset-specific tuning was performed on the EGNN, this out-of-the-box performance indicates that ElemeNet may serve as a competitive baseline for inorganic property prediction with minimal user effort.

**Table 2.** tmQMg property prediction performance for all considered model architectures. All values are test-set MAEs on the splits of Kneiding et al.[84] EGNN* denotes a single multitask model trained on all three targets simultaneously. Each model was trained for up to 1,000 epochs with early stopping. The best-performing ElemeNet model for each task is bolded.

| encoder | readout | ensemble size | HOMO-LUMO gap (hartree) | dipole (debye) | polarizability ($bohr^3$) |
|---|---|---|---|---|---|
| 2D GNN | MLP | 1 | 0.0103 | 1.86 | 9.36 |
| 2D GNN | MLP | 16 | 0.0109 | 1.87 | 8.99 |
| 2D GNN | TF | 1 | 0.0104 | 1.80 | 8.39 |
| 2D GNN | TF | 16 | 0.0118 | 2.04 | 9.49 |
| EGNN | MLP | 1 | **$8.37\times10^{-3}$** | **0.97** | 5.59 |
| EGNN | MLP | 16 | 0.0101 | 1.22 | 8.76 |
| EGNN | TF | 1 | 0.0100 | 1.23 | **5.34** |
| EGNN | TF | 16 | 0.0123 | 1.60 | 6.86 |
| EGNN* | TF | 16 | 0.0120 | 1.78 | 8.05 |
| tmQMg (3D) | | | $7.72\times10^{-3}$–0.0126 (0.0100 average) | 0.71–1.45 (1.10 average) | 4.83–6.81 (5.62 average) |

To evaluate ElemeNet on tasks other than molecule-level property prediction, we benchmark against the pydentate dataset used to train models for predicting metal–ligand coordination.[85] While the original dataset included only graph-level (coordination number) and node-level (coordinating atoms) targets, we extend the dataset here to include additional targets by computing all bond lengths (i.e., an edge-level target) and metal–ligand bond lengths (i.e., a subgraph-level target) for each entry. Because the published pydentate models are 2D, and the inclusion of 3D information already specifies bond lengths, we restrict the comparison to 2D GNN encoders. With these 2D GNNs, we evaluate both MLP and transformer readouts with and without shallow ensembles (Table 3). For coordinating atom and coordination number classification, the performance of our ElemeNet models is comparable to the published pydentate models (Table 3, Figure 6). The transformer readout improved the balanced accuracy on coordinating atom prediction from 92.4% to 96.1%, approaching the pydentate value of 96.9%. Similarly, the model trained with MLP readout exhibits an 84.5% accuracy in coordination number prediction, while the use of attention via the transformer readout improves performance to 86.6%, within 1.9% of the pydentate baseline (Table 3). For bond length and metal–ligand bond length regression, no baselines from the original pydentate work are available. ElemeNet's achieves an MAE of 0.0206 Å on all bond lengths (i.e., an edge-level prediction task) and 0.0505 Å on metal–ligand bond lengths (i.e., a subgraph-level prediction tasks), establishing reference values for these previously unbenchmarked targets (Table 3). The use of shallow ensembles in both regression and classification tasks is observed to have a small impact on performance, reaffirming prior observations on the benefit of increased hyperparameter trials when training with ensemble-based architectures (Figure 6).

**Table 3.** Pydentate dataset performance for the 2D GNN encoder with MLP and TF readouts across node-, graph-, edge-, and subgraph-level targets. Coordinating atom predictions report both raw and class-balanced accuracy. Bond length and metal–ligand bond length results (MAE) were not reported in the original pydentate work and constitute new benchmarks. All values are test-set metrics on the splits of Toney et al.[85] Hyperparameters are optimized for 25 trials of 100 epochs each. Each production model is trained for up to 1,000 epochs with early stopping. The best-performing ElemeNet model for each task is bolded.

| **encoder** | **readout** | **ensemble size** | **coord. atoms (% accuracy/balanced)** | **coord. number (% accuracy)** | **bond lengths (angstroms)** | **metal–ligand bond lengths (angstroms)** |
|---|---|---|---|---|---|---|
| 2D GNN | MLP | 1 | 97.6/92.4 | 84.5 | 0.0208 | 0.0604 |

| 2D GNN | MLP | 16 | 97.6/92.0 | **86.6** | 0.0211 | 0.0607 |
|---|---|---|---|---|---|---|
| 2D GNN | TF | 1 | **98.5/96.1** | 86.6 | **0.0206** | **0.0505** |
| 2D GNN | TF | 16 | 98.4/95.7 | 82.7 | 0.0210 | 0.0512 |
| pydentate (2D) | | | 98.8/96.9 | 88.5 | --- | --- |

Finally, to assess ElemeNet at scale and on biological chemistry, we train on the GEMS dataset of 2.7 million biomolecular fragments with geometries sampled from molecular dynamics trajectories and with DFT-calculated properties.[86] Existing published work on GEMS reports machine-learning force fields rather than direct property prediction, so no comparable literature baseline is available.[86,120] We therefore present ElemeNet results to demonstrate that our workflow scales to multi-million-entry datasets. Motivated by the geometric sensitivity of the targets and the efficacy of attention mechanisms observed on the QM9, tmQMg, and pydentate baselines, we restrict our analysis to the EGNN encoder with transformer readout and compare models trained with and without ensembling. Hyperparameter optimization is impractical at this size with the compute resources available, so default hyperparameters are used throughout.

ElemeNet trains stably on GEMS and yields competitive performance across tasks. The range in model performance spans 0.091–0.119 eV MAE in energy and 0.12–0.13 debye MAE in dipole magnitude (Table 4, Figure 6). Relative atomization energies are not reported in the initial GEMS dataset but are computed here for all structures using the workflow developed in previous work[50] (Supporting Information Text S6). For this quantity, we achieve MAEs on the order of $1.0\times10^{-3}$ eV/(mol g). Shallow ensembling is observed to moderately improve performance across all three prediction tasks, underscoring our prior observations that, under a fixed architecture and set of hyperparameters, ensemble-based training generally improves both predictive accuracy and uncertainty calibration (Table 3 and Figures 3–4). These results highlight that our workflows scale to datasets typical of foundation model training without modification.

**Table 4.** GEMS property prediction performance for the EGNN encoder with transformer readout, with and without shallow ensembling. Default hyperparameters were used due to the computational cost of hyperparameter tuning on this dataset scale (2.7M molecules). Each model was trained for up to 2,000 epochs with early stopping. The best-performing ElemeNet model for each task is bolded.

| encoder | readout | ensemble size | electronic energy (eV) | dipole (debye) | relative atomization energy (eV/mol g) |
|---|---|---|---|---|---|
| EGNN | TF | 1 | 0.119 | 0.13 | $3.75\times10^{-3}$ |
| EGNN | TF | 16 | **0.091** | **0.12** | $\mathbf{3.61\times10^{-3}}$ |

We additionally perform an ablation study on charge and spin to determine the utility of the explicit charge and spin encoding supported natively in ElemeNet. Due to the prevalence of charged species in the overall dataset (i.e., 36% of the 2.7M total entries), we expect GEMS to be an excellent test case indicating the utility of our dedicated charge-spin neural network. We identify all charged structures in the GEMS test set and run inference on each model separately with the true charges. We compare the resulting summary statistics to the output of running inference with all entries intentionally mislabeled as neutral (Supporting Information Table S19). The resulting increased MAE in all cases indicates the utility of our explicit embedding scheme.

Taken together, these benchmarks demonstrate that ElemeNet produces performant property prediction models across organic, inorganic, coordination, and biological chemistry and across graph-, node-, edge-, and subgraph-level targets using a single unified workflow. While production deployments would benefit from longer training, larger batch sizes, and more rigorous hyperparameter optimization, the results presented here are obtained with deliberately modest compute budgets and establish that ElemeNet provides competitive baselines with minimal user effort.

**4. Conclusions**

In summary, we developed the ElemeNet software package for training and using advanced machine learning (ML) models with support for molecules with elemental compositions across the periodic table. This software was designed to make ML model training accessible for researchers in the chemical and physical sciences without extensive training in computation and ML. We designed ElemeNet to support common 2D and 3D molecular representations, including SMILES, mol2, and xyz formats, and define initial representations compatible with elements 1–100. This enables the development of ML models for systems that defy traditional covalent bonding descriptions. Optional graph-level encodings we developed enable the training of models which predict properties as explicit functions of molecular charge and spin. The code was built to support regression and classification targets in both single- and multitask contexts, enabling multiscale predictions on atomic (node), bond (edge), molecular (graph), and our newly introduced moiety (subgraph) targets. State-of-the-art E(3) equivariant and attention mechanisms were included through EGNN and transformer architectures, in addition to 2D GNN and MLP models. Uncertainty quantification by shallow ensembles and latent space distance were made available for all model architectures. Our codebase is parallelizable across multiple GPUs. We showed through benchmarking studies on representative datasets from organic, inorganic, coordination, and biological chemistry that ElemeNet models exhibit strong performance even at the scale of millions of molecules. Across these broad tests, ElemeNet either matches or exceeds nearly all literature baselines. We anticipate that ElemeNet will reduce the barrier to entry in ML for chemistry, democratizing the training and use of advanced machine learning models across the periodic table.

## AUTHOR INFORMATION

**Corresponding Author**

*email:hjkulik@mit.edu

**Author Contributions**

CRediT: Jacob W. Toney: Conceptualization, Data curation, Formal Analysis, Investigation, Methodology, Software, Validation, Visualization, Writing – original draft, Writing – review & editing; Samir Darouich: Conceptualization, Data curation, Formal Analysis, Investigation, Methodology, Software, Validation, Writing – review & editing; Yiran Wang: Investigation, Software, Validation; Aaron G. Garrison: Investigation, Software; Johannes Kästner: Funding acquisition, Supervision; Heather J. Kulik: Conceptualization, Funding acquisition, Project administration, Resources, Supervision, Writing – review & editing.

**Notes**

The authors declare no competing financial interest.

## ASSOCIATED CONTENT

**Supporting Information**. Initial node representations used by encoder; initial edge representations used by encoder; bond interpretation scheme; charge and spin determination scheme; charge and spin embedding scheme; optional bulk features used in initial representations; optional xTB features used in initial representations; full space of hyperparameters considered; performance and calibration of QM9 atomization energy model ensembles; performance and calibration of QM9 gap model ensembles; performance and calibration of tmQMg dipole model ensembles; performance and calibration of tmQMg gap model ensembles; binary classification ensemble reduction to binary cross-entropy; multiclass classification ensemble reduction to multiclass cross-entropy; list of unparsable SMILES in the pydentate dataset; performance and calibration of binary pydentate model ensembles; performance and calibration of multiclass pydentate model ensembles; optional RDKit features used in initial representations; relevant functions, keywords, and commands defined within ElemeNet; restricted space of hyperparameters considered during "quicksearch"; evaluation metrics available in ElemeNet; uncertainty metrics available in ElemeNet; results of 3D GNN models reported in original tmQMg work; procedure used to calculate atomization energies for the GEMS dataset; charge-spin encoding ablation study. (PDF)

This material is available free of charge via the Internet at http://pubs.acs.org.

**Data and Software Availability Statement**

The ElemeNet codebase is available publicly via GitHub at https://github.com/hjkgrp/ElemeNet.

All data required to reproduce this work is provided either in the Supporting Information PDF file or in the Zenodo repository.[121]

ACKNOWLEDGMENT

Funding was primarily provided by a UPI from The Dow Chemical Company. J.W.T. was partially supported by a Leslye Miller Fraser and Darryl M. Fraser Fellowship from the MIT School of Engineering. H.J.K. is supported by a Simon Family Faculty Research Innovation Fund and an Alfred P. Sloan Fellowship in Chemistry. S.D. was partially funded by the Ministry of Science, Research and the Arts Baden-Wuerttemberg in the Artificial Intelligence Software Academy (AISA) and the Deutsche Forschungsgemeinschaft (DFG, German Research Foundation) EXC2075 – 390740016 under Germany's Excellence Strategy. S.D. thanks IMPRS-IS (International Max Planck Research School for Intelligent Systems) for the support. The authors acknowledge the MIT SuperCloud and Lincoln Laboratory Supercomputing Center and MIT Office of Research Computing and Data for providing HPC resources that have contributed to the research results reported in this work. The authors thank Adam H. Steeves and Mathias Niepert for providing a critical reading of the manuscript. Finally, the authors thank Heecheol

Jang, Akash K. Ball, Changhwan Oh, Sebastian G. Pujet, and Teya S. Bergamaschi for valuable technical discussions and scientific insights.

**For Table of Contents Use Only**

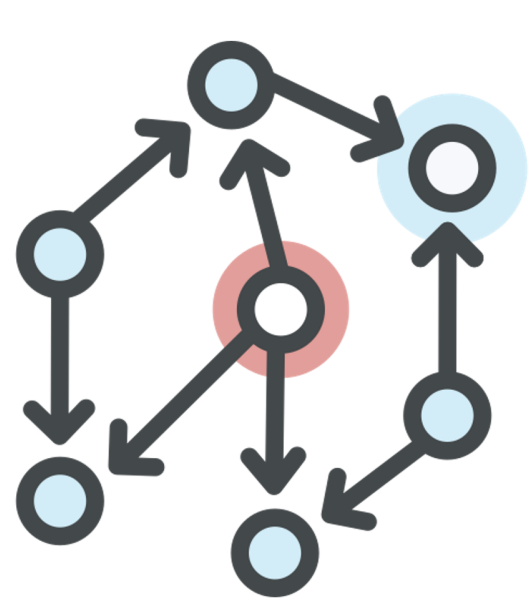

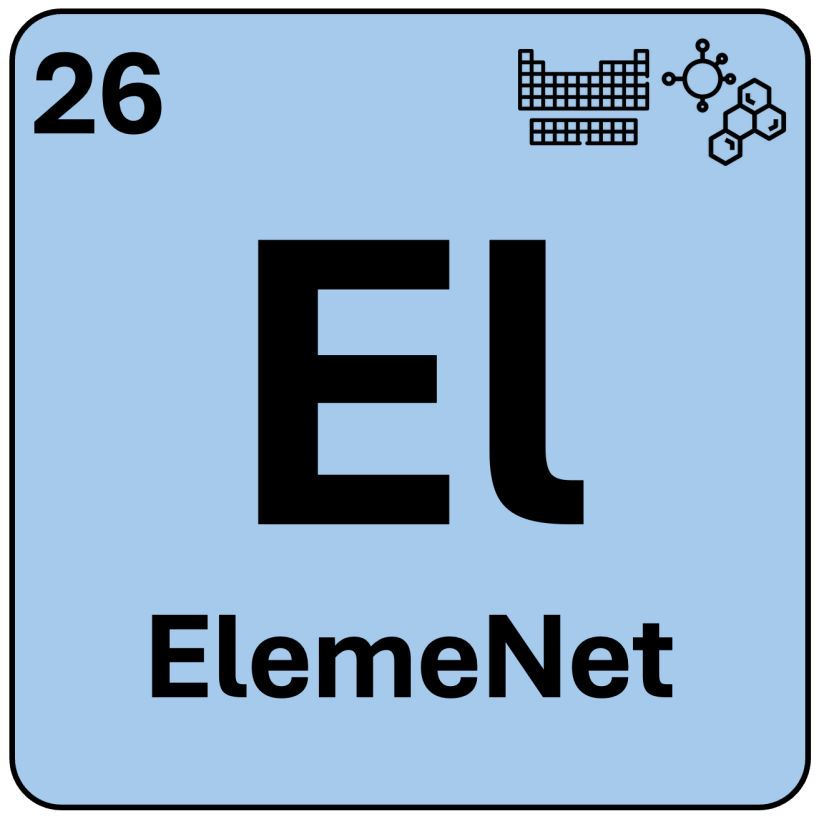

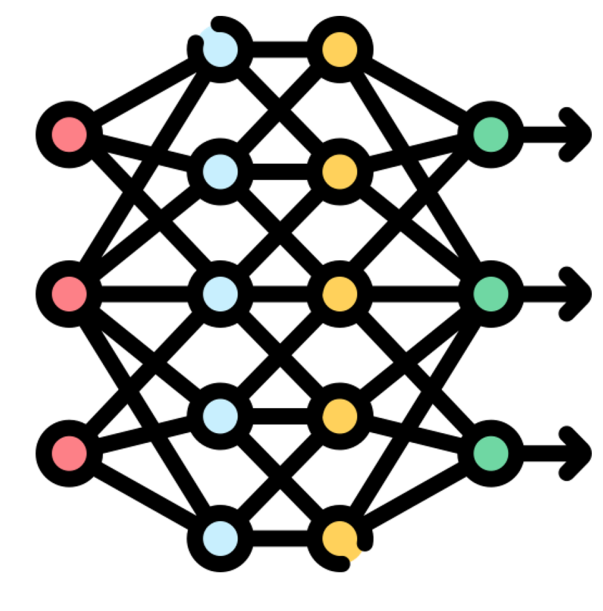

**Supporting Information for**

***ElemeNet: Multiscale Molecular Machine Learning with Uncertainty Quantification Across the Periodic Table***

Jacob W. Toney[1,2,#], Samir Darouich[1,3,4,#], Yiran Wang[1,2], Aaron G. Garrison[1], Johannes Kästner[3], and Heather J. Kulik[1,2,5,*]

[1]*Department of Chemical Engineering, Massachusetts Institute of Technology, Cambridge, MA 02139, USA*

[2]*Center for Computational Science and Engineering, Massachusetts Institute of Technology, Cambridge, MA 02139, USA*

[3]*Institute for Theoretical Chemistry, University of Stuttgart, Germany*

[4]*Institute for Artificial Intelligence, University of Stuttgart, Germany*

[5]*Department of Chemistry, Massachusetts Institute of Technology, Cambridge, MA 02139, USA*

#These authors contributed equally

*corresponding author email: hjkulik@mit.edu

**Contents**

**Table S1.** Initial node representations used by the encoder. Atomic mass, covalent radius, Pauling electronegativity, period, and group are obtained from WebElements[1]. van der Waals radii are from Alvarez[2], while polarizability values are from Schwerdtfeger and Nagle[3]. Valence electrons are reported for the $Ns$, $Np$, $(N-1)d$, and $(N-2)f$ orbitals, where $N$ is period, and are extracted from the NIST Atomic Spectra Database[4] and https://periodictable.com[5]. Experimentally-measured first ionization potentials and electron affinities are extracted from mendeleev[6] v0.19.0. All of these values may be found in the `globalvars.py` file of the ElemeNet codebase, and the user may add missing elements or adjust values accordingly by editing this file.

| feature (units) | dimension | range | encoding |
|---|---|---|---|
| atomic mass (amu) | 1 | ["H": 1.0079, "Fm": 257] | min-max scaling |
| covalent radius (Å) | 1 | ["H": 0.37, "Fr": 3.48] | min-max scaling |
| van der Waals radius (Å) | 1 | ["H": 1.2, "Cs": 3.48] | min-max scaling |
| electronegativity | 1 | ["He": 0, "F": 3.98] | min-max scaling |
| polarizability (bohr$^3$) | 1 | ["He": 1.38375, "Cs": 400.9] | min-max scaling |
| ionization potential (eV) | 1 | ["Cs": 3.89, "He": 24.59] | min-max scaling |
| electron affinity (eV) | 1 | ["He": -19.7, "Cl": 3.61] | min-max scaling |
| coordination number | 1 | [0, 12] | min-max scaling |
| valence electrons in *s* orbitals | 3 | [0, 3] | one-hot |
| valence electrons in *p* orbitals | 7 | [0, 6] | one-hot |
| valence electrons in d orbitals | 11 | [0, 10] | one-hot |
| valence electrons in *f* orbitals | 15 | [0, 14] | one-hot |
| period | 7 | [1, 7] | one-hot |
| group | 18 | [1, 18] | one-hot |
| number of hydrogens | 1 | [0, 8] | min-max scaling |

**Table S2.** Initial edge representations used by the encoder. Bond type may be either single, double, triple, or aromatic. "Has metal". Estimated bond length is equal to the sum of covalent radii between the bonded atoms. For `edge_invariant=True`, only estimate bond length and "has metal" features are used. EGNN encoder uses distance-based embedding in lieu of these edge features.

| feature (units) | dimension | encoding |
|---|---|---|
| bond type | 4 | one-hot |
| estimated bond length (Å) | 1 | integer |
| has metal | 1 | integer |

**Text S1.** Bond interpretation scheme.

ElemeNet supports the training of 2D graph neural networks (GNNs) based on molecular connectivity for cif, cml, mol, mol2, sdf, SMILES, and xyz formats. Most of these formats explicitly encode connectivity, with the exception of cif and xyz formats, which only include 3D coordinates. For these inputs, bonds are automatically interpreted based on the pairwise distances between all atoms in the input structure. If the distance is within some threshold ($\tau$), a bond is added; otherwise, the atoms remain unbonded. ElemeNet uses the sum of covalent radii ($r_{\text{cov}}$) for each pair of atoms ($i, j$) as its bonding threshold multiplied by a constant scale-factor ($\alpha$):

$$\tau = \alpha\left(r_{\text{cov},i} + r_{\text{cov},j}\right) \tag{S1}$$

This scale-factor is set to 1.0 by default but may be adjusted by the user through the `bond_scale_factor` keyword.

We note that this approach only encodes the presence of a bond, not bond order (i.e., all interpreted bonds for cif and xyz inputs are defined as single bonds). This assumption has little effect, since initial molecular representations are updated by message passing regardless so long as a bond of any type exists between two atoms. We also note that our E(3) equivariant GNN architectures do not require any bonding information, as they rely exclusively on pairwise distances between points.

**Text S2.** Charge and spin determination scheme.

ElemeNet explicitly considers the charge and spin of all molecules. These values are determined automatically, but the code allows for user intervention. Charge is determined by assuming all molecules are neutral, while allowed spin multiplicities are determined by counting the total number of electrons in the system (assuming all atoms are neutral) and adding on the determined charge. If the total number of electrons is even, the molecule is assumed to be a singlet, while a doublet spin state is assumed is the total number of electrons is odd (i.e., we assume a low-spin state by default).

These assumptions may be relaxed by several mechanisms. First, the user may optionally define the charge and/or spin multiplicity of each entry in their dataset by including the columns `charge` and `spinmult` in their input .csv files and specifying these via the `feature_columns` argument to `training_pipeline`. If only charge is provided, the assumed spin multiplicity is updated accordingly by subtracting the specified charge from the calculated total number of electrons. Similarly, if only spin is provided, the charge is still assumed to be neutral, and the `spinmult` argument indicates whether the molecule is in a higher-spin state. Internal consistency is still required (e.g., specifying a spin multiplicity of 2 for a complex with an even number of electrons will raise an error), but this may be optionally overwritten with the `charge_spin_override` keyword.

ElemeNet also supports the specification of atomic partial charges via an `atomic_charges` column in the input .csv file. These atom-centered charges are concatenated to initial node representations, and are also used to inform the net charge of a molecule. If molecular charge is not specified, it is calculated as the sum of atomic charges. If both `atomic_charges` and `charge` are provided, internal consistency is again confirmed (i.e., charge must equal the sum of atomic charges).

**Text S3.** Charge and spin embedding scheme.

Charge and spin information is automatically encoded for all molecules (Text S2). By default, these graph-level attributes are concatenated to the learned node representations, after pooling if applicable (i.e., for graph-level targets). For increased expressivity, ElemeNet also supports projecting charge and spin through a small one-layer neural network, with the resulting learned graph embeddings concatenated to (pooled) node representations. The hidden layer size of this charge-spin neural network is controlled by the `graph_attr_hidden_dim` keyword and is tuned during hyperparameter optimization over a search space of 8, 16, 32, or 64 neurons. In the case that charge and spin provides no valuable information (e.g., in datasets such as QM9 which include only neutral, closed-shell molecules), this network also supports a hidden dimension size of "None", corresponding to standard concatenation of charge and spin to node representations. We note that while our above discussion is limited to charge and spin (i.e., the only graph-level features defined for all molecules by default), if additional graph-level descriptors are provided, these attributes will be projected and concatenated in the same manner.

**Table S3.** Optional bulk features used in initial representations. All bulk features are defined on a per-node basis. These features may be optionally concatenated to initial node representations through the `use_bulk` keyword (default is False). All values are obtained from https://periodictable.com[5]. These values are all stored in the `globalvars.py` file and may be readily modified by the user.

| feature (units) | dimension | range | encoding |
|---|---|---|---|
| melting temperature (K) | 1 | ["He": 0, "C": 3823.15] | min-max scaling |
| boiling temperature (K) | 1 | ["At": 0, "Re": 5869.15] | min-max scaling |
| heat of fusion (kJ/mol) | 1 | ["Pu": 0, "C": 105] | min-max scaling |
| heat of vaporization (kJ/mol) | 1 | ["Am": 0, "W": 800] | min-max scaling |
| specific heat capacity (J/kg K) | 1 | ["Pm": 0, "H": 14300] | min-max scaling |
| thermal conductivity (W/m K) | 1 | ["Po": 0, "Ag": 430] | min-max scaling |
| electrical conductivity (S/m) | 1 | ["H": 0, "Ag": 6.2E7] | min-max scaling |
| lattice angles | 3 | $\alpha$: [0, $\pi/2$]<br>$\beta$: [0, 2.313085]<br>$\gamma$: [0, $2\pi/3$] | unscaled |
| lattice constants | 3 | a: [0, 11.45]<br>b: [0, 12.845]<br>c: [0, 26.25] | unscaled |
| phase | 3 | {"solid", "liquid", "gas"} | one-hot |
| crystal structure | 12 | {"bcc", "bcm", "bco", "bct", "cub", "fcc", "fco", "hex", "mono", "ortho", "tet", "tri"} | one-hot |
| space group | 19 | {"2", "11", "12", "14", "15", "62", "63", "64", "70", "139", "141", "152", "166", "194", "217", "221", "225", "227", "229"} | one-hot |

**Table S4.** Optional xTB features used in initial representations. These features may be optionally concatenated to initial node, edge, and graph representations through the `use_xtb` keyword (default is "False"). The results are directly obtained from the output of a GFN2-xTB[7] calculation initialized at the user-provided input geometry and the determined charge and spin state (Text S2). As xTB features require 3D information as inputs, they are not supported for SMILES representations, which only encode connectivity. Dimension refers to the feature size per-unit of scope. For example, the per-node dimension of the "atomic charge" feature is 1, but there are a total of $N$ atomic charges reported for a system with $N$ atoms.

| **feature (units)** | **scope** | **dimension** | **encoding** |
|---|---|---|---|
| energy (Ha) | graph | 1 | unscaled |
| dipole (e bohr) | graph | 1 | unscaled |
| gap (Ha) | graph | 1 | unscaled |
| atomic charge (e) | node | 1 | unscaled |
| atomic energy (Ha) | node | 1 | unscaled |
| bond order (e) | edge | 1 | unscaled |

**Table S5.** Full space of hyperparameters considered.

| hyperparameter | search space | description | default |
|---|---|---|---|
| learning rate | [1e-6, 1e-2] | Initial learning rate used by optimizer. Search performed in continuous log-space. | 1e-4 |
| weight decay | [1e-6, 1e-1] | Weight decay regularization used by optimizer. Search performed in continuous log-space. | 0 |
| encoder layers | {3, 4, 5, 6, 7, 8, 9} | Number of encoder layers. | 3 |
| encoder neurons | {64, 128, 256, 512} | Neurons in first encoder hidden layer. | 128 |
| encoder shape | {"constant", "increasing", "decreasing", "hourglass", "pyramid"} | Shape of encoder network. Used by GNN encoder only. EGNN encoder uses "constant" shape. | "constant" |
| encoder dropout | [0, 0.4] | Dropout rate among encoder neurons. Search performed in continuous space. | 0.1 |
| encoder activation | {"relu", "tanh", "leakyrelu", "gelu", "silu"} | Encoder activation functions. Used by GNN encoder only. EGNN encoder uses separate equivariant activations. | "relu" |
| encoder convolution | {"gcnconv", "graphconv", "gineconv", "nnconv", "gat"} | Encoder convolution operators. Used by GNN encoder only. EGNN encoder uses separate equivariant convolutions. | "graphconv" |
| encoder pooling | {"mean", "sum", "max"} | Function for aggregating node to graph representations. Only required for graph or subgraph-level targets. | "mean" |
| EGNN inverse sublayers | {1, 2, 3} | Number of equivariant sub-layers used per encoder hidden layer for coordinate updates. Used by EGNN encoder only. | 2 |
| EGNN attention | {True, False} | Whether to use attention in equivariant message-passing. Used by EGNN encoder only. | False |
| EGNN distance embedding | {True, False} | Whether to project distances to higher dimensional space via Gaussian basis functions. Used by EGNN encoder only. | True |
| EGNN Gaussians | {32, 64, 128} | Number of Gaussian basis functions to use for distance embedding projection. Used by EGNN encoder only. | 64 |
| EGNN aggregation | {"mean", "sum"} | Aggregation function for equivariant message-passing. Used by EGNN encoder only. | "sum" |
| readout layers | {1, 2, 3, 4, 5} | Number of readout layers. | 3 |
| readout neurons | {64, 128, 256, 512} | Neurons in first readout hidden layer. | 128 |
| readout shape | {"constant", "increasing", "decreasing", "hourglass", "pyramid"} | Shape of readout network. Used by MLP readout only. Transformer readout uses "constant" shape. | "constant" |
| readout dropout | [0, 0.4] | Dropout rate among readout neurons. Search performed in continuous space. | 0.1 |
| readout activation | {"relu", "tanh", "leakyrelu", "gelu", "silu"} | Readout activation functions. | "relu" |
| readout norm | {True, False} | Whether to normalize activations in readout hidden layers. | True |
| graph attribute hidden size | {8, 16, 32, 64, None} | Hidden dimension to which graph attributes are projected. None results in no change to original graph attribute size. | None |
| TF attention heads | {2, 4, 8} | Number of TF attention heads. Used by TF readout only. | 8 |
| TF expansion | {2, 4, 8} | Projection of TF attention heads. Used by TF readout only. | 4 |

**Table S6.** Performance and calibration of QM9 atomization energy ensemble models. All models trained utilize a 2D GNN encoder and either an MLP or transformer (TF) readout architecture with default hyperparameters. Models are trained for 500 epochs with early stopping. Data splits of QM9 from Heid et al. are used.[8] The hyperparameter-optimized models originally trained by Heid et al. achieve an MAE of 1.11 kcal/mol on atomization energy.[8] Our models are not hyperparameter-optimized and are not fully trained to validation loss saturation. Spearman rank correlation ($\rho$) measures the extent to which ensemble variance or latent space distance (LSD) between $N$ nearest neighbors correlates with predictive error, with higher values indicating better calibration. Bold indicates the best result for a given metric, and the arrow indicates what direction corresponds to a better value for a metric. Wall time is reported for end-to-end runs (i.e., including preprocessing, training, and inference).

| model | ensemble size | wall time (s) | MAE (kcal/mol) | ensemble variance $\rho$ (↑) | LSD, $N = 1$ $\rho$ (↑) | LSD, $N = 10$ $\rho$ (↑) | LSD, $N = 200$ $\rho$ (↑) |
|---|---|---|---|---|---|---|---|
| GNN-MLP | 1 | 6543 | 10.01 | --- | -0.069 | -0.067 | -0.122 |
| | 2 | 7041 | 11.09 | 0.203 | 0.088 | 0.114 | 0.103 |
| | 4 | 7032 | 10.38 | 0.354 | 0.002 | 0.023 | 0.084 |
| | 8 | 7050 | **7.68** | 0.379 | **0.111** | **0.153** | **0.218** |
| | 16 | 6895 | 8.93 | 0.387 | 0.047 | 0.063 | 0.104 |
| | 32 | 7088 | 9.06 | 0.397 | 0.055 | 0.082 | 0.097 |
| | 64 | 7059 | 9.32 | **0.462** | 0.097 | 0.111 | 0.106 |
| GNN-TF | 1 | 10943 | 1.61 | --- | 0.184 | 0.245 | 0.246 |
| | 2 | 11224 | 5.17 | 0.379 | 0.083 | 0.120 | 0.138 |
| | 4 | 11138 | **1.45** | 0.391 | 0.237 | 0.307 | 0.309 |
| | 8 | 11228 | 1.58 | 0.392 | 0.203 | 0.269 | 0.277 |
| | 16 | 11156 | 1.47 | 0.372 | **0.250** | **0.325** | **0.319** |
| | 32 | 11217 | 1.46 | 0.382 | 0.226 | 0.289 | 0.289 |
| | 64 | 10878 | 1.46 | **0.397** | 0.229 | 0.310 | 0.316 |

**Table S7.** Performance and calibration of QM9 gap ensemble models. All models trained utilize a 2D GNN encoder and either an MLP or transformer (TF) readout architecture with default hyperparameters. Models are trained for 500 epochs with early stopping. Data splits of QM9 from Heid et al. are used.[8] The hyperparameter-optimized models originally trained by Heid et al. achieve an MAE of $3.14 \times 10^{-3}$ hartree on HOMO-LUMO gap.[8] Our models are not hyperparameter optimized and are not fully trained to validation loss saturation. Spearman rank correlation ($\rho$) measures the extent to which ensemble variance or latent space distance (LSD) between $N$ nearest neighbors correlates with predictive error, with higher values indicating better calibration. Bold indicates the best result for a given metric, and the arrow indicates what direction corresponds to a better value for a metric. Wall time is reported for end-to-end runs (i.e., including preprocessing, training, and inference).

| model | ensemble size | wall time (s) | MAE (hartree) | ensemble variance $\rho$ (↑) | LSD, $N = 1$ $\rho$ (↑) | LSD, $N = 10$ $\rho$ (↑) | LSD, $N = 200$ $\rho$ (↑) |
|---|---|---|---|---|---|---|---|
| GNN-MLP | 1 | 6737 | 5.80e-3 | --- | 0.098 | 0.076 | 0.049 |
| | 2 | 7328 | 5.84e-3 | 0.408 | **0.238** | 0.225 | 0.166 |
| | 4 | 7112 | 5.56e-3 | 0.415 | 0.195 | 0.187 | 0.138 |
| | 8 | 7213 | 5.64e-3 | 0.421 | 0.219 | 0.207 | 0.158 |
| | 16 | 7069 | 5.66e-3 | **0.426** | 0.223 | 0.218 | 0.168 |
| | 32 | 6900 | **5.53e-3** | 0.423 | 0.234 | **0.234** | **0.185** |
| | 64 | 6909 | 5.62e-3 | 0.411 | 0.220 | 0.215 | 0.165 |
| GNN-TF | 1 | 10888 | **4.48e-3** | --- | **0.289** | **0.318** | 0.230 |
| | 2 | 10972 | 5.10e-3 | 0.384 | 0.266 | 0.291 | 0.250 |
| | 4 | 11166 | 5.35e-3 | 0.386 | 0.247 | 0.286 | 0.247 |
| | 8 | 11237 | 5.28e-3 | 0.383 | 0.245 | 0.278 | 0.243 |
| | 16 | 10706 | 5.47e-3 | 0.397 | 0.253 | 0.298 | 0.275 |
| | 32 | 10740 | 5.18e-3 | **0.403** | 0.273 | 0.314 | **0.288** |
| | 64 | 11050 | 5.44e-3 | 0.393 | 0.249 | 0.285 | 0.266 |

**Table S8.** Performance and calibration of tmQMg dipole ensemble models. All models trained utilize an EGNN encoder and either an MLP or transformer (TF) readout architecture with default hyperparameters. Models are trained for 500 epochs with early stopping. Data splits of tmQMg from Kneiding et al. are used.[9] Hyperparameter-optimized models originally trained by Kneiding et al. without DFT inputs achieve an average MAE of 1.10 debye on dipole.[9] Our models are not hyperparameter-optimized and are not fully trained to validation loss saturation. Spearman rank correlation ($\rho$) measures the extent to which ensemble variance or latent space distance (LSD) between $N$ nearest neighbors correlates with predictive error, with higher values indicating better calibration. Bold indicates the best result for a given metric, and the arrow indicates what direction corresponds to a better value for a metric. Wall time is reported for end-to-end runs (i.e., including preprocessing, training, and inference).

| model | ensemble size | wall time (s) | MAE (debye) | ensemble variance $\rho$ (↑) | LSD, $N = 1$ $\rho$ (↑) | LSD, $N = 10$ $\rho$ (↑) | LSD, $N = 200$ $\rho$ (↑) |
|---|---|---|---|---|---|---|---|
| EGNN-MLP | 1 | 5044 | **1.14** | --- | 0.268 | 0.2524 | 0.2104 |
| | 2 | 5084 | 1.66 | 0.378 | 0.283 | 0.3043 | 0.2752 |
| | 4 | 5077 | 1.27 | 0.470 | 0.362 | 0.3757 | 0.3485 |
| | 8 | 5077 | 1.25 | **0.480** | 0.366 | 0.3851 | 0.3616 |
| | 16 | 5089 | 1.41 | 0.480 | **0.370** | **0.3932** | **0.3663** |
| | 32 | 5643 | 1.50 | 0.470 | 0.330 | 0.3345 | 0.2909 |
| | 64 | 5621 | 1.39 | 0.464 | 0.314 | 0.3277 | 0.2897 |
| EGNN-TF | 1 | 5770 | **1.39** | --- | 0.341 | 0.3347 | 0.2733 |
| | 2 | 6261 | 1.68 | 0.390 | 0.230 | 0.2741 | 0.2756 |
| | 4 | 6251 | 1.51 | 0.428 | 0.322 | 0.3475 | 0.3326 |
| | 8 | 5636 | 1.55 | 0.423 | 0.354 | 0.3931 | 0.3835 |
| | 16 | 6262 | 1.61 | **0.444** | **0.372** | 0.4045 | 0.3899 |
| | 32 | 6269 | 1.63 | 0.438 | 0.333 | 0.3818 | 0.3979 |
| | 64 | 6251 | 1.64 | 0.436 | 0.365 | **0.4124** | **0.4180** |

**Table S9.** Performance and calibration of tmQMg gap ensemble models. All models trained utilize an EGNN encoder and either an MLP or transformer (TF) readout architecture with default hyperparameters. Models are trained for 500 epochs with early stopping. Data splits of tmQMg from Kneiding et al. are used.[9] Hyperparameter-optimized models originally trained by Kneiding et al. without DFT inputs achieve an average MAE of $1.00\times10^{-2}$ hartree on HOMO-LUMO gap.[9] Our models are not hyperparameter-optimized and are not fully trained to validation loss saturation. Spearman rank correlation ($\rho$) measures the extent to which ensemble variance or latent space distance (LSD) between $N$ nearest neighbors correlates with predictive error, with higher values indicating better calibration. Bold indicates the best result for a given metric, and the arrow indicates what direction corresponds to a better value for a metric. Wall time is reported for end-to-end runs (i.e., including preprocessing, training, and inference).

| model | ensemble size | wall time (s) | MAE (hartree) | ensemble variance $\rho$ (↑) | LSD, $N = 1$ $\rho$ (↑) | LSD, $N = 10$ $\rho$ (↑) | LSD, $N = 200$ $\rho$ (↑) |
|---|---|---|---|---|---|---|---|
| EGNN-MLP | 1 | 5093 | 0.0099 | --- | 0.158 | 0.126 | 0.087 |
| | 2 | 5466 | 0.0105 | 0.301 | 0.231 | 0.224 | 0.187 |
| | 4 | 5462 | **0.0098** | **0.330** | **0.276** | **0.274** | **0.247** |
| | 8 | 5202 | 0.0099 | 0.315 | 0.273 | 0.269 | 0.227 |
| | 16 | 5161 | 0.0099 | 0.319 | 0.262 | 0.257 | 0.217 |
| | 32 | 5468 | 0.0101 | 0.306 | 0.227 | 0.218 | 0.179 |
| | 64 | 5465 | 0.0103 | 0.311 | 0.246 | 0.238 | 0.199 |
| EGNN-TF | 1 | 5578 | **0.0107** | --- | **0.346** | **0.299** | 0.156 |
| | 2 | 6054 | 0.0112 | **0.330** | 0.243 | 0.261 | 0.231 |
| | 4 | 5651 | 0.0123 | 0.271 | 0.199 | 0.199 | 0.143 |
| | 8 | 5644 | 0.0120 | 0.279 | 0.221 | 0.214 | 0.160 |
| | 16 | 6058 | 0.0123 | 0.289 | 0.237 | 0.235 | 0.188 |
| | 32 | 5705 | 0.0122 | 0.312 | 0.260 | 0.272 | **0.234** |
| | 64 | 5611 | 0.0124 | 0.272 | 0.228 | 0.244 | 0.215 |

**Text S4.** Proof that the negative log-likelihood of an ensemble of Bernoulli predictions under a Beta prior is exactly equivalent to binary cross-entropy.

For binary classification, we may place a Beta prior on binary class probability $p$ and model observations $y$ as draws from a Bernoulli distribution conditional on $p$:

$$p \sim \text{Beta}(\alpha, \beta), \qquad y|p \sim \text{Bernoulli}(p) \tag{S2}$$

where $\alpha$ and $\beta$ are the shape parameters defining the Beta distribution. The joint probability distribution $P(y, p|\alpha, \beta)$ may be factored as:

$$P(y, p|\alpha, \beta) = P(y|p)P(p|\alpha, \beta) \tag{S3}$$

where $P(y|p) = p^y(1-p)^{1-y}$ is the known likelihood of a Bernoulli distribution. The Beta prior density is $P(p|\alpha, \beta) = p^{\alpha-1}(1-p)^{\beta-1}/B(\alpha, \beta)$, where $B(\alpha, \beta)$ is the Beta function.

We may integrate with respect to $p$ to obtain $P(y|\alpha, \beta)$, the marginal log-likelihood of $y$ given $\alpha$ and $\beta$:

$$P(y|\alpha, \beta) = \int_0^1 p^y(1-p)^{1-y} \cdot \frac{p^{\alpha-1}(1-p)^{\beta-1}}{B(\alpha, \beta)} dp \tag{S4}$$

$$= \frac{1}{B(\alpha, \beta)} \int_0^1 p^{\alpha+y-1}(1-p)^{\beta-y}\, dp \tag{S5}$$

$$= \frac{B(\alpha + y, \beta + 1 - y)}{B(\alpha, \beta)} \tag{S6}$$

The negative log-likelihood (NLL) is:

$$\mathcal{L}_{\text{NLL}} = -\log \frac{B(\alpha + y, \beta + 1 - y)}{B(\alpha, \beta)} \tag{S7}$$

$$= \log B(\alpha, \beta) - \log B(\alpha + y, \beta + 1 - y) \tag{S8}$$

Because a binary class observation $y$ must be either 0 or 1, the marginal log-likelihood may exhibit only two cases. Considering both cases and rewriting the Beta function explicitly in terms of Gamma functions as $B(\alpha, \beta) = \Gamma(\alpha)\Gamma(\beta)/\Gamma(\alpha + \beta)$:

**Case** $y = 1$:

$$P(1|\alpha, \beta) = \frac{B(\alpha + 1, \beta)}{B(\alpha, \beta)} \tag{S9}$$

$$= \frac{\Gamma(\alpha + 1)\Gamma(\beta)}{\Gamma(\alpha + \beta + 1)} \cdot \frac{\Gamma(\alpha + \beta)}{\Gamma(\alpha)\Gamma(\beta)} \tag{S10}$$

$$= \frac{\alpha\Gamma(\alpha)}{\Gamma(\alpha)} \cdot \frac{\Gamma(\alpha+\beta)}{(\alpha+\beta)\Gamma(\alpha+\beta)} \tag{S11}$$

$$= \frac{\alpha}{\alpha+\beta} = \bar{p} \tag{S12}$$

**Case** $y = 0$:

$$P(0|\alpha,\beta) = \frac{B(\alpha,\beta+1)}{B(\alpha,\beta)} \tag{S13}$$

$$= \frac{\Gamma(\alpha)\Gamma(\beta+1)}{\Gamma(\alpha+\beta+1)} \cdot \frac{\Gamma(\alpha+\beta)}{\Gamma(\alpha)\Gamma(\beta)} \tag{S14}$$

$$= \frac{\beta\Gamma(\beta)}{\Gamma(\beta)} \cdot \frac{\Gamma(\alpha+\beta)}{(\alpha+\beta)\Gamma(\alpha+\beta)} \tag{S15}$$

$$= \frac{\beta}{\alpha+\beta} = 1 - \bar{p} \tag{S16}$$

utilizing the relations $\Gamma(x+1) = x\Gamma(x)$ and the definitions of Beta distribution shape parameters, where $\bar{p}$ is the mean probability.

Combining both cases and taking the negative log-likelihood results in:

$$P(y|\alpha,\beta) = \bar{p}^{y}(1-\bar{p})^{1-y} \tag{S17}$$

$$\mathcal{L}_{\mathrm{NLL}} = -\,\mathrm{y}\log\bar{p} - (1-y)\log(1-\bar{p}) = \mathrm{BCE}(\bar{p}, y) \tag{S18}$$

Therefore, the negative log-likelihood of the Beta-Bernoulli is equivalent to binary cross-entropy (BCE).

**Text S5.** Proof that the negative log-likelihood of an ensemble of Categorical predictions under a Dirichlet prior is exactly equivalent to multiclass cross-entropy.

Let $K$ be the number of classes and $y \in \{1, \cdots, K\}$ be the true class label for an observation. For multiclass classification, we place a Dirichlet prior on per-class probabilities $\mathbf{p}$ and model observations $y$ as draws from a Categorical distribution conditional on $\mathbf{p}$:

$$\mathbf{p} \sim \text{Dirichlet}(\boldsymbol{\alpha}), \qquad y|\mathbf{p} \sim \text{Categorical}(\mathbf{p}) \tag{S19}$$

where $\boldsymbol{\alpha} = \{\alpha_1, \cdots, \alpha_K\}$ are concentration parameters defining the Dirichlet distribution. The total concentration is defined as $\alpha_0 = \sum_{k=1}^{K} \alpha_k$. The joint probability distribution $P(y, \mathbf{p}|\boldsymbol{\alpha})$ factors as:

$$P(y, \mathbf{p}|\boldsymbol{\alpha}) = P(y|\mathbf{p})P(\mathbf{p}|\boldsymbol{\alpha}) \tag{S20}$$

where $P(y|\mathbf{p}) = \prod_{k=1}^{K} p_k^{\delta_{yk}}$ is the likelihood of a Categorical distribution, $p_k$ is the probability of class $k$, and $\delta_{yk}$ is the Kronecker delta. The Dirichlet prior density is $P(\mathbf{p}|\boldsymbol{\alpha}) = \frac{1}{B(\boldsymbol{\alpha})} \prod_{k=1}^{K} p_k^{\alpha_k - 1}$, where $B(\boldsymbol{\alpha})$ is the multivariate Beta function and $\alpha_k$ is the concentration parameter of dimension $k$.

Integrating $\mathbf{p}$ gives $P(y|\boldsymbol{\alpha})$, the marginal log-likelihood of $y$ given $\boldsymbol{\alpha}$:

$$P(y|\boldsymbol{\alpha}) = \int_{\Delta^{K-1}} P(y|\mathbf{p})P(\mathbf{p}|\boldsymbol{\alpha})d\mathbf{p} \tag{S21}$$

$$= \int_{\Delta^{K-1}} \prod_{k=1}^{K} p_k^{\delta_{yk}} \frac{1}{B(\boldsymbol{\alpha})} \prod_{k=1}^{K} p_k^{\alpha_k - 1} \, d\mathbf{p} \tag{S22}$$

$$= \frac{1}{B(\boldsymbol{\alpha})} \int_{\Delta^{K-1}} \prod_{k=1}^{K} p_k^{\delta_{yk} + \alpha_k - 1} \, d\mathbf{p} \tag{S23}$$

where $\Delta^{K-1}$ is the standard $K-1$ simplex support upon which a $K$-dimensional Dirichlet distribution is defined. The integrand is proportional to a Dirichlet density with concentration parameters $\boldsymbol{\alpha}' = \{\alpha_1, \cdots, \alpha_y + 1, \cdots, \alpha_K\}$, where all values are equivalent to the $\boldsymbol{\alpha}$ case aside from the $y$-th which is incremented by one due to the $\delta_{yk} - 1$ term. Because the integral over the support of any probability density equals one, the integral must equal the normalizing constant of this offset Dirichlet:

$$\int_{\Delta^{K-1}} \prod_{k=1}^{K} p_k^{\alpha'_k - 1} \, d\mathbf{p} = B(\boldsymbol{\alpha}') \tag{S24}$$

Resulting in the simplified marginal log-likelihood expression:

$$P(y|\boldsymbol{\alpha}) = \frac{B(\boldsymbol{\alpha}')}{B(\boldsymbol{\alpha})} \tag{S25}$$

We may simplify the expression by rewriting the multivariate Beta function in terms of Gamma functions (Eq. S25) and invoking the Gamma recurrence relation, $\Gamma(x+1) = x\Gamma(x)$:

$$B(\boldsymbol{\alpha}) = \frac{1}{\Gamma(\alpha_0)} \prod_{k=1}^{K} \Gamma(\alpha_k) \tag{S26}$$

Concretely:

$$P(y|\boldsymbol{\alpha}) = \frac{B(\boldsymbol{\alpha}')}{B(\boldsymbol{\alpha})} = \frac{\Gamma(\alpha_0)}{\Gamma(\alpha_0')} \prod_{k=1}^{K} \frac{\Gamma(\alpha_k')}{\Gamma(\alpha_k)} \tag{S27}$$

Because $\alpha_k = \alpha_k'$ for all $k$ components besides the $y$-th element, $B(\boldsymbol{\alpha})$ and $B(\boldsymbol{\alpha}')$ are equivalent aside from the $y$-th concentration parameter. The total concentrations also differ only by one, i.e., $\alpha_0' = \alpha_0 + 1$. Therefore:

$$P(y|\boldsymbol{\alpha}) = \frac{\Gamma(\alpha_0)}{\Gamma(\alpha_0 + 1)} \prod_{k=1}^{K} \frac{\Gamma(\alpha_k')}{\Gamma(\alpha_k)} \tag{S28}$$

$$= \frac{\Gamma(\alpha_0)}{\Gamma(\alpha_0 + 1)} \cdot \frac{\Gamma(\alpha_y')}{\Gamma(\alpha_y)} \tag{S29}$$

$$= \frac{\Gamma(\alpha_0)}{\Gamma(\alpha_0 + 1)} \cdot \frac{\Gamma(\alpha_y + 1)}{\Gamma(\alpha_y)} \tag{S30}$$

$$= \frac{\Gamma(\alpha_0)}{\alpha_0 \Gamma(\alpha_0)} \cdot \frac{\alpha_y \Gamma(\alpha_y)}{\Gamma(\alpha_y)} \tag{S31}$$

$$P(y|\boldsymbol{\alpha}) = \frac{\alpha_y}{\alpha_0} = \mathbb{E}[P(y|\mathbf{p})] = p_y \tag{S32}$$

where $\mathbb{E}[P(y|\mathbf{p})]$ is the prior mean of the $y$-th component under a Dirichlet. In the case of an ensemble of predictors, $\mathbb{E}[P(y|\mathbf{p})]$ is the ensemble mean prediction for class $y$, expressed as $p_y$. Therefore, the marginal probability of observing class $y$ under a Dirichlet-Categorical model is equivalent to the prior mean for class $y$, identical to the binary case (Text S4).

Multiclass cross-entropy collapses to a single term for supervised classification on $\mathbf{y}$, a target vector which is equal to one for the true class index ($y$) and zero elsewhere, resulting in only the $-\log(p_y)$ term remaining. Thus, multiclass cross-entropy is equivalent to an ensemble under the Dirichlet-Categorical NLL, equivalent to binary cross-entropy and Beta-Bernoulli NLL (Text S4).

**Table S10.** Summary table of parsable and unparsable SMILES in the pydentate dataset. While all 66,355 molecules were successfully parsed in the original models trained by Toney et al.[10], a small subset of 66 SMILES (i.e., less than 0.1% of the dataset) were not parsable by our ElemeNet models, which we attribute to differences in RDKit versions and sanitization procedures used. Specifically, the original pydentate models were trained with RDKit version 2023.09.4 without explicit sanitization (`sanitize=False`), while this work uses version 2025.03.5 and applies the sanitization routine built-in by RDKit (`sanitize=True`).[11] We report the differences in training, validation, and test set size here and provide the missing SMILES in the associated Zenodo repository.[12] Only ElemeNet models trained on the pydentate dataset and using SMILES string representations are affected (i.e., those trained to predict coordinating atoms and coordination number). ElemeNet models trained on the pydentate dataset using mol2 representations (i.e., those trained to predict bond lengths and metal–ligand bond lengths) use the full dataset.

| data split | full dataset | after invalid SMILES removed |
|---|---|---|
| train | 53,009 | 52,957 |
| validation | 6,730 | 6,722 |
| test | 6,616 | 6,610 |
| total | 66,355 | 66,289 |

**Table S11.** Performance and calibration of binary pydentate ensemble models. All models trained utilize a 2D GNN encoder and either an MLP or transformer (TF) readout architecture with default hyperparameters. Models are trained for 500 epochs with early stopping. The target property is coordinating atom identity (i.e., a binary node-level target). Training, validation, and test splits are adapted from the pydentate dataset from Toney et al.[10], omitting any ligands with invalid SMILES (Table S10). Hyperparameter-optimized models trained on the full dataset by Toney et al. achieve an accuracy of 98.8% and AUROC of 1.00.[10] Our models are not hyperparameter-optimized and are not fully trained to validation loss saturation. Separate in-distribution (train/val/test) and out-of-distribution (OOD) sets are created by holding out all ligands with phosphorus coordinating atoms as a separate OOD set. Expected calibration error (ECE) measures the extent to which model confidence in predicted probabilities correlates with accuracy, with lower values indicating better calibration. Bold indicates the best result for a given metric, and the arrow indicates what direction corresponds to a better value for a metric. Wall time is reported for end-to-end runs (i.e., including preprocessing, training, and inference).

| model | ensemble size | wall time (s) | in-distribution | | | out-of-distribution | | |
|---|---|---|---|---|---|---|---|---|
| | | | accuracy (↑) | AUROC (↑) | ECE (↓) | accuracy (↑) | AUROC (↑) | ECE (↓) |
| GNN-MLP | 1 | 3948 | 0.972 | 0.985 | 0.0021 | 0.942 | 0.531 | 0.0429 |
| | 2 | 4054 | 0.972 | 0.985 | 0.0023 | 0.940 | 0.551 | **0.0414** |
| | 4 | 4016 | 0.972 | 0.985 | 0.0021 | 0.941 | **0.570** | 0.0426 |
| | 8 | 3997 | 0.972 | 0.985 | 0.0020 | **0.943** | 0.552 | 0.0422 |
| | 16 | 4095 | 0.972 | 0.985 | 0.0025 | 0.942 | 0.534 | 0.0435 |
| | 32 | 3919 | 0.972 | 0.985 | 0.0017 | 0.941 | 0.473 | 0.0426 |
| | 64 | 3905 | 0.972 | 0.985 | 0.0016 | 0.942 | 0.529 | 0.0421 |
| GNN-TF | 1 | 7755 | 0.982 | 0.996 | 0.0028 | 0.909 | 0.596 | 0.0668 |
| | 2 | 7824 | 0.982 | 0.996 | 0.0027 | 0.907 | **0.697** | 0.0618 |
| | 4 | 7781 | 0.981 | 0.996 | 0.0028 | 0.915 | 0.679 | 0.0702 |
| | 8 | 7884 | 0.982 | 0.996 | 0.0028 | 0.910 | 0.660 | 0.0664 |
| | 16 | 7841 | 0.982 | 0.996 | **0.0011** | **0.924** | 0.694 | **0.0556** |
| | 32 | 7797 | 0.982 | 0.996 | 0.0028 | 0.911 | 0.686 | 0.0569 |
| | 64 | 7837 | 0.982 | 0.996 | 0.0025 | 0.918 | 0.630 | 0.0785 |

**Table S12.** Performance and calibration of multiclass pydentate model ensembles. All models trained utilize a 2D GNN encoder and either an MLP or transformer readout architecture with default hyperparameters. Models trained for 500 epochs with early stopping. The target property is coordination number (i.e., a multiclass graph-level target). Training, validation, and test splits are adapted from the pydentate dataset from Toney et al.[10], omitting any ligands with invalid SMILES (Table S10). Hyperparameter-optimized models originally trained on the full dataset by Toney et al. achieve an accuracy of 88.5% and AUROC of 0.98.[10] Our models are not hyperparameter optimized and are not fully trained to validation loss saturation. Separate in-distribution (train/val/test) and out-of-distribution (OOD) sets are created by holding out all ligands with phosphorus coordinating atoms as a separate OOD set. F1 score is macro-averaged across classes. Expected calibration error (ECE) measures the extent to which model confidence in predicted probabilities correlates with accuracy, with lower values indicating better calibration. Wall time is reported for end-to-end runs (i.e., including preprocessing, training, and inference).

| model | ensemble size | wall time (s) | in-distribution | | | out-of-distribution | | |
|---|---|---|---|---|---|---|---|---|
| | | | accuracy (↑) | F1 (↑) | ECE (↓) | accuracy (↑) | F1 (↑) | ECE (↓) |
| GNN-MLP | 1 | 3186 | 0.836 | 0.793 | 0.0501 | **0.481** | **0.308** | 0.3402 |
| | 2 | 3278 | 0.839 | 0.793 | 0.0499 | 0.411 | 0.251 | 0.4030 |
| | 4 | 3227 | 0.846 | **0.807** | **0.0398** | 0.401 | 0.265 | 0.4267 |
| | 8 | 3199 | **0.848** | 0.806 | 0.0453 | 0.428 | 0.268 | 0.3934 |
| | 16 | 3322 | 0.842 | 0.788 | 0.0416 | 0.451 | 0.300 | **0.3276** |
| | 32 | 3307 | 0.841 | 0.801 | 0.0402 | 0.442 | 0.280 | 0.3500 |
| | 64 | 3246 | 0.840 | 0.793 | 0.0454 | 0.422 | 0.254 | 0.3740 |
| GNN-TF | 1 | 7016 | 0.754 | 0.704 | 0.0686 | 0.490 | 0.317 | 0.2345 |
| | 2 | 7153 | 0.745 | 0.699 | 0.0320 | 0.505 | 0.338 | 0.1781 |
| | 4 | 7132 | **0.762** | **0.726** | 0.0412 | 0.453 | 0.284 | 0.2487 |
| | 8 | 7190 | 0.758 | 0.712 | 0.0490 | 0.511 | **0.361** | 0.2071 |
| | 16 | 7097 | 0.759 | 0.715 | **0.0275** | 0.446 | 0.297 | 0.2402 |
| | 32 | 7190 | 0.757 | 0.709 | 0.0461 | 0.480 | 0.324 | 0.2280 |
| | 64 | 7211 | 0.758 | 0.705 | 0.0295 | **0.515** | 0.347 | **0.1555** |

**Table S13.** Optional RDKit features used in initial representations. These features may be optionally concatenated to initial node and edge representations through the `rdkit_features` keyword (default is "False"). Features are directly obtained from RDKit parsing of SMILES strings, and are not supported for any other representations (e.g., mol2, xyz). These RDKit features are also best defined for small organic molecules, and are not recommended for use with more complex chemistry (e.g., organometallics). Dimension refers to the feature size per-unit of scope. For example, the per-node dimension of the "formal charge" feature is 1, but there are a total of $N$ formal charges reported for a system with $N$ atoms.

| feature (units) | scope | dimension | range | encoding |
|---|---|---|---|---|
| formal charge (e) | node | 1 | {-2, -1, 0, 1, 2} | unscaled |
| hybridization | node | 1 | {"$sp$", "$sp^2$", "$sp^3$", "$sp^3d$", "$sp^3d^2$"} | unscaled |
| chiral tag | node | 1 | {0, 1, 2, 3} | unscaled |
| aromaticity | node | 1 | {0, 1} | unscaled |
| conjugation | edge | 1 | {0, 1} | unscaled |
| cyclic | edge | 1 | {0, 1} | unscaled |
| stereochemistry | edge | 1 | {0, 1, 2, 3, 4, 5} | unscaled |

**Table S14.** Relevant functions, keywords, and commands defined within ElemeNet.

| utility (path) | ElemeNet syntax | description |
|---|---|---|
| preprocessing (src/elemenet/preprocess.py) | atomic_fingerprint | Featurize nodes with atomic properties. |
| | bond_fingerprint | Featurize edges with bonds properties. |
| | mol_fingerprint | Featurize graphs with molecular properties. |
| | feature_columns | Specify extra user-provided features at the node, edge, or graph level. |
| | charge | Keyword to specify molecular charge via feature_columns. |
| | spinmult | Keyword to specify molecular spin multiplicity via feature_columns. |
| | mol_column | Column containing molecular graph inputs. |
| | graph_format | Specify molecular graph type. Must be either cif, cml, mol, mol2, sdf, smiles, or xyz. |
| | target_column | Column containing target property. |
| | center_column | Column containing atom indices around which to generate subgraphs for moiety predictions (optional). |
| | k_hops | Number of atoms away from center atom to consider for moiety predictions (optional). |
| | group_by | Column upon which to stratify molecules when preparing data splits (optional). |
| encoder architectures (src/elemenet/encoder.py) | GNN_Encoder | 2D graph neural network (GNN) encoder architecture. |
| | EGNN_Encoder | Equivariant 3D GNN encoder architecture. |
| readout architectures (src/elemenet/readout.py) | MLP | Multilayer perceptron readout architecture. |
| | Transformer | Transformer readout architecture. |
| model training (src/elemenet/trainer.py) | LOSS_FN | Loss function for model training. Must be either mse, mae, cross_entropy, bce, ensemble_regression, ensemble_binary_classification, or ensemble_multi_classification. |
| | OPTIMIZER | Optimizer for model training. Must be either adam, sgd, rmsprop, adagrad, or adamw. |
| | SCHEDULER | Learning rate scheduler for model training (optional). Must be either cosine, plateau, or onecycle. |
| hyperparameter search space (src/elemenet/hypersearch.py) | search_space | Defines global search space for hyperparameter optimization. |
| | search_space_encoder | Defines encoder search space for hyperparameter optimization. |
| | search_space_readout | Defines readout search space for hyperparameter optimization. |
| training wrapper (src/elemenet/training.py) | training_pipeline | Wrapper function for end-to-end model training, including preprocessing and inference. |
| inference wrapper (src/elemenet/inference.py) | inference_pipeline | Wrapper function for end-to-end inference from trained model and new dataset. |
| command-line interface (src/scripts) | elemenet_train | Command-line interface (CLI) for training_pipeline. |
| | elemenet_inference | CLI for inference_pipeline. |

**Table S15.** Restricted space of hyperparameters considered "quicksearch".

| hyperparameter | search space | description | default |
|---|---|---|---|
| learning rate | [1e-5, 1e-2] | Initial learning rate used by optimizer. Search performed in continuous log-space. | 1e-4 |
| weight decay | [1e-5, 1e-1] | Weight decay regularization used by optimizer. Search performed in continuous log-space. | 0 |
| encoder layers | {3, 5, 7} | Number of encoder layers. | 3 |
| encoder neurons | {128, 256} | Neurons in first encoder hidden layer. | 128 |
| encoder dropout | [0, 0.3] | Dropout rate among encoder neurons. Search performed in continuous space. | 0.1 |
| encoder convolution | {"gcnconv", "graphconv", "gineconv", "nnconv", "gat"} | Encoder convolution operators. Used by GNN encoder only. EGNN encoder uses separate equivariant convolutions. | "graphconv" |
| readout layers | {1, 2, 3} | Number of readout layers. | 3 |
| readout neurons | {128, 256} | Neurons in first readout hidden layer. | 128 |
| readout dropout | [0, 0.3] | Dropout rate among readout neurons. Search performed in continuous space. | 0.1 |
| TF attention heads | {2, 4, 8} | Number of TF attention heads. Used by TF readout only. | 8 |
| TF expansion | {2, 4, 8} | Projection of TF attention heads. Used by TF readout only. | 4 |

**Table S16.** Evaluation metrics available in ElemeNet. All evaluation metrics are supported for all scopes (i.e., node, edge, graph, and subgraph).

| metric | supported tasks | theoretical bounds | description |
|---|---|---|---|
| MAE | regression | $[0, \infty)$ | Mean absolute error |
| MSE | regression | $[0, \infty)$ | Mean squared error |
| RMSE | regression | $[0, \infty)$ | Root mean squared error |
| accuracy | classification | [0, 1] | Overall rate of correct predictions.<br>May be weighted by class frequency. |
| precision | classification | [0, 1] | Rate of correct positive predictions over all positive predictions.<br>May be weighted by class frequency. |
| recall | classification | [0, 1] | Rate of correct positive predictions over true positive and false negative predictions.<br>May be weighted by class frequency. |
| F1 | classification | [0, 1] | Harmonic mean of precision and recall. May be weighted by class frequency. |
| ROC-AUC | classification | [0, 1] | Receiver operating characteristic area under curve. Measures ability to distinguish between classes.<br>Averaged across classes for multiclass datasets. |

**Table S17.** Uncertainty metrics available in ElemeNet. Latent space distances and cosine similarity are only supported for graph-level targets, as saving node and edge-level latent space embeddings is computationally intractable (i.e., results in massive $N \times H$ matrices, for $N$ nodes or edges and $H$ hidden dimension). Confidence and entropy metrics are only defined for classification tasks. Only ensemble standard deviations are defined for all tasks (i.e., regression and classification) and scopes (i.e., node, edge, graph, and subgraph).

| metric | supported tasks | supported scopes | theoretical bounds | description |
|---|---|---|---|---|
| latent space distance | regression, classification | graph | [0, ∞) | Latent space distance to closest $N$ data points in training set. |
| cosine similarity | regression, classification | graph | [-1, 1] | Cosine of angle formed by the vectors pointing from the origin to the test point and closest point in training data. |
| confidence | classification | node, edge, graph, subgraph | binary: [0, 1]<br>K-class: [0, 1/K] | Minimum distance of a prediction from 0 or 1. |
| normalized confidence | classification | node, edge, graph, subgraph | [0, 1] | Minimum distance of a prediction from 0 or 1, normalized for number of classes. |
| entropy | classification | node, edge, graph, subgraph | [0, log(K)) | Shannon entropy, defined as expected value of information content. |
| normalized entropy | classification | node, edge, graph, subgraph | [0, 1) | Shannon entropy, defined as expected value of information content, normalized for number of classes. |
| ensemble standard deviation | regression, classification | node, edge, graph, subgraph | [0, ∞) | Standard deviation of predictions across ensemble heads. |

**Table S18.** Results of 3D GNN models reported in original tmQMg work. All values are directly taken as reported in Kneiding et al.[9] Models requiring quantum mechanical calculations as inputs (i.e., "NatQG") are not considered.

| tmQMg baseline architecture | HOMO-LUMO gap (hartree) | dipole (debye) | polarizability (bohr$^3$) |
|---|---|---|---|
| MXMNet | $9.36\times10^{-3}$ | 0.943 | 4.83 |
| SchNet | 0.0126 | 1.45 | 6.81 |
| EdgeUpdate | 0.0102 | 1.13 | 5.67 |
| DimeNet++ | 0.0103 | 1.28 | 5.37 |
| ALIGNN | $7.72\times10^{-3}$ | 0.705 | 5.43 |
| average | 0.0100 | 1.10 | 5.62 |

**Text S6.** Procedure used to calculate atomization energies for the GEMS dataset.

To compute atomization energies, we performed single-point energy calculations on all unique atom types (i.e., H, C, N, O, S) and charges (i.e., -3, -2, -1, 0, 1, 2, 3) in the GEMS dataset.[13] For each atom type and charge combination, the first five accessible spin states were considered. For example, O (charge -1) was modeled as a doublet, quartet, sextet, octet, and decet, while N (charge 3) was only considered in the singlet, triplet, and quintet states. These results were used to determine the ground state spin for each combination of atom and charge. For a given structure with reported electronic energy, the difference in electronic energy and single-atom energies was used to calculate atomization energy. Charged species were handled using the procedure outlined in Garrison et al., with the net charge being satisfied by adding or removing electrons from the atoms with the smallest difference in energy upon addition/removal.[14] As atomization energy is an extensive property which scales with size, we normalize all properties by molecular mass to yield a relative atomization energy with units of eV/(mol g). All calculations used the PBE0 functional[15] and def2-TZVPP basis set[16,17] for consistency with the reported GEMS electronic energies.[13] All calculations utilized Psi4 version 1.9.1.[18] All atom energies may be found in the Zenodo repository.[12]

**Table S19.** Charge-spin encoding ablation study. Test set of GEMS dataset is filtered to contain only charged species. Charge is either defined explicitly (i.e., explicit charge True) or intentionally mislabeled as neutral (i.e., explicit charge False). Inference is performed on each set using existing trained models.

| encoder | readout | ensemble size | explicit charge | electronic energy (eV) | dipole (debye) | relative atomization energy (eV/mol g) |
|---|---|---|---|---|---|---|
| EGNN | TF | 1 | True | 0.127 | 0.155 | $3.82\times10^{-3}$ |
| EGNN | TF | 1 | False | 3.057 | 0.158 | 0.297 |
| EGNN | TF | 16 | True | 0.096 | 0.150 | $3.92\times10^{-3}$ |
| EGNN | TF | 16 | False | 0.629 | 0.151 | $2.92\times10^{-2}$ |